\documentstyle[12pt]{article}

\catcode`\@=11
\def\href#1#2{#2}
\def\@citex[#1]#2{%
\if@filesw \immediate \write \@auxout {\string \citation {#2}}\fi
\@tempcntb\m@ne \let\@h@ld\relax \def\@citea{}%
\@cite{%
  \@for \@citeb:=#2\do {%
    \@ifundefined {b@\@citeb}%
      {\@h@ld\@citea\@tempcntb\m@ne{\bf ?}%
      \@warning {Citation `\@citeb ' on page \thepage \space undefined}}%
      {\@tempcnta\@tempcntb \advance\@tempcnta\@ne%
      \@tempcntb\number\csname b@\@citeb \endcsname \relax%
      \ifnum\@tempcnta=\@tempcntb 
        \ifx\@h@ld\relax%
          \edef \@h@ld{\@citea\csname b@\@citeb\endcsname}%
        \else%
          \edef\@h@ld{\ifmmode{-}\else--\fi\csname b@\@citeb\endcsname}%
        \fi%
      \else
        \@h@ld\@citea\csname b@\@citeb \endcsname%
        \let\@h@ld\relax%
      \fi}%
    \def\@citea{,\penalty\@highpenalty\,}%
  }\@h@ld
}{#1}}

\def\@citeb#1#2{{[#1]\if@tempswa , #2\fi}}
\def\@citeu#1#2{{$^{#1}$\if@tempswa , #2\fi }}
\def\@citep#1#2{{#1\if@tempswa , #2\fi}}

\def\bcites{         
        \catcode`\@=11
        \let\@cite=\@citeb
        \catcode`\@=12
}

\def\upcites{         
        \catcode`\@=11
        \let\@cite=\@citeu
        \catcode`\@=12
}

\def\plaincites{      
        \catcode`\@=11
        \let\@cite=\@citep
        \catcode`\@=12
}

\newcount\hour
\newcount\minute
\newtoks\amorpm
\hour=\time\divide\hour by 60
\minute=\time{\multiply\hour by 60 \global\advance\minute by-\hour}
\edef\standardtime{{\ifnum\hour<12 \global\amorpm={am}%
        \else\global\amorpm={pm}\advance\hour by-12 \fi
        \ifnum\hour=0 \hour=12 \fi
        \number\hour:\ifnum\minute<10 0\fi\number\minute\the\amorpm}}
\edef\militarytime{\number\hour:\ifnum\minute<10 0\fi\number\minute}

\def\draftlabel#1{{\@bsphack\if@filesw {\let\thepage\relax
   \xdef\@gtempa{\write\@auxout{\string
      \newlabel{#1}{{\@currentlabel}{\thepage}}}}}\@gtempa
   \if@nobreak \ifvmode\nobreak\fi\fi\fi\@esphack}
        \gdef\@eqnlabel{#1}}
\def\@eqnlabel{}
\def\@vacuum{}
\def\marginnote#1{}
\def\draftmarginnote#1{\marginpar{\raggedright\scriptsize\tt#1}}
\def\draft{
        \pagestyle{plain}
        \overfullrule=2pt
        \oddsidemargin -.5truein
        \def\@oddhead{\sl \phantom{\today\quad\militarytime} \hfil
        \smash{\Large\sl DRAFT} \hfil \today\quad\militarytime}
        \let\@evenhead\@oddhead
        \let\label=\draftlabel
        \let\marginnote=\draftmarginnote
        \def\ps@empty{\let\@mkboth\@gobbletwo
        \def\@oddfoot{\hfil \smash{\Large\sl DRAFT} \hfil}
        \let\@evenfoot\@oddhead}
        \def\@eqnnum{(\theequation)\rlap{\kern\marginparsep\tt\@eqnlabel}%
        \global\let\@eqnlabel\@vacuum}  }

\def\blackfonts{
        \font\blackboard=msbm10 scaled\magstep1
        \font\blackboards=msbm8
        \font\blackboardss=msbm6
}

\def\nblack{            
        \def\ZZ{{Z \n{10} Z}}
        \def\NN{{N \n{14} N}}
        \def\CC{{C \n{11} C}}
        \def\RR{{R \n{11} R}}
        \def\QQ{{Q \n{12} Q}}
        \def\PP{{P \n{11} P}}
}

\def\prep{         
        \catcode`\@=11
        \input art10.sty
        \catcode`\@=12
        
        \let\small\null
        \def\blackfonts{
                \font\blackboard=msbm10
                \font\blackboards=msbm7
                \font\blackboardss=msbm5
        }
        \let\sl\it
        \twocolumn
        \sloppy
        \voffset=-2.54truecm
        \hoffset=-2.54truecm
        \flushbottom
        \parindent 1em
        \leftmargini 2em
        \leftmarginv .5em
        \leftmarginvi .5em
        \marginparwidth 48pt
        \marginparsep 10pt
        \setlength{\columnsep}{2truecm}
        \setlength{\textwidth}{25.4truecm}
        \setlength{\textheight}{17truecm}
        \baselineskip=16pt
        \oddsidemargin .18truein
        \evensidemargin .17truein
}

\def\eqalign#1{\null\,\vcenter{\openup\jot\m@th
  \ialign{\strut\hfil$\displaystyle{##}$&$\displaystyle{{}##}$\hfil
      \crcr#1\crcr}}\,}
\def\eqalignno#1{\displ@y \tabskip\centering
  \halign to\displaywidth{\hfil$\@lign\displaystyle{##}$\tabskip\z@skip
    &$\@lign\displaystyle{{}##}$\hfil\tabskip\centering
    &\llap{$\@lign##$}\tabskip\z@skip\crcr
    #1\crcr}}

\def\section{\@startsection {section}{1}{\z@}{3.ex plus 1ex minus
 .2ex}{2.ex plus .2ex}{\large\bf}}
\def\subsection{\@startsection{subsection}{2}{\z@}{2.75ex plus 1ex minus
 .2ex}{1.5ex plus .2ex}{\bf}}

\def\appendix{{\newpage\section*{Appendices}}\let\appendix\section%
        {\setcounter{section}{0}
        \gdef\thesection{\Alph{section}}}\section}

\def\abstract{\if@twocolumn
\section*{Abstract}
\else 
\begin{center}
{\bf Abstract\vspace{-.5em}\vspace{0pt}}
\end{center}
\quotation
\fi}

\catcode`\@=12

\def\noj#1,#2,{{\bf #1} (19#2)\ }
\def\jou#1,#2,#3,{{\sl #1\/ }{\bf #2} (19#3)\ }
\def\ann#1,#2,{{\sl Ann.\ Physics\/ }{\bf #1} (19#2)\ }
\def\cmp#1,#2,{{\sl Comm.\ Math.\ Phys.\/ }{\bf #1} (19#2)\ }
\def\cq#1,#2,{{\sl Class.\ Quantum Grav.\/ }{\bf #1} (19#2)\ }
\def\cqg#1,#2,{{\sl Class.\ Quantum Grav.\/ }{\bf #1} (19#2)\ }
\def\ijmp#1,#2,{{\sl Int.\ J.\ Mod.\ Phys.\/ }{\bf A#1} (19#2)\ }
\def\jmp#1,#2,{{\sl J.\ Math.\ Phys.\/ }{\bf #1} (19#2)\ }
\def\grg#1,#2,{{\sl Gen.\ Rel.\ Grav.\/ }{\bf #1} (19#2)\ }
\def\mpl#1,#2,{{\sl Mod.\ Phys.\ Lett.\/ }{\bf A#1} (19#2)\ }
\def\nc#1,#2,{{\sl Nuovo Cim.\/ }{\bf #1} (19#2)\ }
\def\np#1,#2,{{\sl Nucl.\ Phys.\/ }{\bf B#1} (19#2)\ }
\def\pl#1,#2,{{\sl Phys.\ Lett.\/ }{\bf #1B} (19#2)\ }
\def\pla#1,#2,{{\sl Phys.\ Lett.\/ }{\bf #1A} (19#2)\ }
\def\pr#1,#2,{{\sl Phys.\ Rev.\/ }{\bf #1} (19#2)\ }
\def\prd#1,#2,{{\sl Phys.\ Rev.\/ }{\bf D#1} (19#2)\ }
\def\prl#1,#2,{{\sl Phys.\ Rev.\ Lett.\/ }{\bf #1} (19#2)\ }
\def\prp#1,#2,{{\sl Phys.\ Rept.\/ }{\bf #1C} (19#2)\ }
\def\ptp#1,#2,{{\sl Prog.\ Theor.\ Phys.\/ }{\bf #1} (19#2)\ }
\def\ptpsup#1,#2,{{\sl Prog.\ Theor.\ Phys.\/ Suppl.\/ }{\bf #1} (19#2)\ }
\def\rmp#1,#2,{{\sl Rev.\ Mod.\ Phys.\/ }{\bf #1} (19#2)\ }
\def\yadfiz#1,#2,#3[#4,#5]{{\sl Yad.\ Fiz.\/ }{\bf #1} (19#2) #3%
\ [{\sl Sov.\ J.\ Nucl.\ Phys.\/ }{\bf #4} (19#2) #5]}
\def\zh#1,#2,#3[#4,#5]{{\sl Zh.\ Exp.\ Theor.\ Fiz.\/ }{\bf #1} (19#2) #3%
\ [{\sl Sov.\ Phys.\ JETP\/ }{\bf #4} (19#2) #5]}

\def\beq{\begin{equation}}
\def\eeq{\end{equation}}
\def\beqar{\begin{eqnarray}}
\def\eeqar{\end{eqnarray}}

\def\nfrac#1#2{{\displaystyle{\vphantom1\smash{\lower.5ex\hbox{\small$#1$}}%
        \over\vphantom1\smash{\raise.25ex\hbox{\small$#2$}}}}}

\def\n#1{\mskip-#1mu}

\def\to{\rightarrow}

\def\lae{\mathrel{\mathop{\smash{\lower .5 ex \hbox{$\stackrel<\sim$}}}}}
\def\lae{\mathrel{\mathop{\smash{\lower .5 ex \hbox{$\stackrel>\sim$}}}}}

\def\l:{\mathopen{:}\,}
\def\r:{\,\mathclose{:}}

\def\[{\left[}          \def\]{\right]}
\def\({\left(}          \def\){\right)}
\def\<{\left<}          \def\>{\right>}

\catcode`\@=11
\def\theequation{\arabic{equation}}
\catcode`\@=12

\nblack
\bcites

\nblack

\catcode`\@=11
\def\theequation{\thesection.\arabic{equation}}
\@addtoreset{equation}{section}
\@addtoreset{footnote}{section}
\@addtoreset{footnote}{subsection}
\catcode`\@=12

\def\half{\nfrac12}

\def\Z{{\bf Z}}
\def\R{{\bf R}}

\newcommand{\beqn}{\begin{equation}}
\newcommand{\eeqn}{\end{equation}}
\newcommand{\beqnarray}{\begin{eqnarray}}
\newcommand{\eeqnarray}{\end{eqnarray}}
\begin{document}

\input psfig

\begin{titlepage}

\begin{center}
November, 1996
\hfill       IASSNS-HEP-96/121 \\
\hfill                  hep-th/9611230

\vskip 1 cm
{\large \bf Type IIB Superstrings,
BPS Monopoles, And Three-Dimensional Gauge Dynamics
\\}
\vskip 0.1 cm
\vskip 0.5 cm
{Amihay Hanany and Edward Witten
\footnote{Research supported in part by NSF Grant PHY-9513835.}
}
\vskip 0.2cm
{\sl
hanany; witten@ias.edu \\
School of Natural Sciences \\
Institute for Advanced Study \\
Olden Lane, Princeton, NJ 08540, USA
}

\end{center}

\vskip 0.5 cm

\begin{abstract}
We propose an explanation via string theory
of the correspondence between the Coulomb
branch of certain
three-dimensional supersymmetric
gauge theories and certain moduli spaces of 
magnetic monopoles. The same construction also gives an explanation,
via $SL(2,\Z)$ duality of Type IIB superstrings, of the
recently discovered ``mirror symmetry" in three dimensions.
New phase transitions in  three 
dimensions as well as new infrared fixed points and even
new coupling constants not present in the known Lagrangians
are predicted from the string theory construction.  
An important role in the construction is played by a novel
aspect of brane dynamics in which a third brane is created
when two branes cross.

\end{abstract}

\end{titlepage}

\section{Introduction}

Recently, there has been much investigation of the dynamics
of three-dimensional supersymmetric gauge theories with
$N=4$ supersymmetry.  Two of the most remarkable discoveries
are as follows.  
\begin{enumerate}
\item  In some cases, the Coulomb branch of vacua
is isomorphic, as a hyper-Kahler manifold, to a moduli space of
three-dimensional monopole solutions in a {\it different} gauge
theory.  For instance, the three-dimensional theory with
$SU(k)$ gauge theory and no hypermultiplets has for its Coulomb
branch the moduli space of $k$-monopole configurations of an $SU(2)$
gauge theory in $3+1$ dimensions.  (This was shown in \cite{SW}
for $k=2$; the  generalization to higher $k$ was proposed in \cite{CH}.)
\item In some cases, there is a ``mirror symmetry'' between two
models  in which the Coulomb branch of one is the Higgs
branch of the other  and vice-versa \cite{SI}.

\end{enumerate}

The purpose of this paper is to give an explanation of these two
phenomena by embedding them in a suitable string theory context.
More specifically, we will describe a simple configuration of
threebranes and fivebranes in ten-dimensional Type IIB superstring
theory from whose basic properties the phenomena mentioned
in the last paragraph can be deduced -- at least for a large class of
models.  In the process, we will encounter a number of unusual
phenomena, including  field theories that can be generalized to include
parameters not seen in the classical Lagrangian as well as things
that are by now more familiar, such as
 exotic phase transitions and infrared fixed points.
Also, in the course of our investigation, we will have to understand
a new feature of brane dynamics in which a third brane is created
when two branes cross.

\section{A Type IIB configuration}

In this paper, we will study certain
supersymmetric configurations of fivebranes
and threebranes in Type IIB superstring theory.
In detail these will be as follows.

We work in ten-dimensional Minkowski space with time coordinate $x^0$ and space
coordinates $x^1,\ldots,x^9$.  Let $Q_L$ and $Q_R$ be the 
supercharges generated by left- and right-moving world-sheet degrees of 
freedom.
The Type IIB theory is chiral; the supercharges obey $\bar \Gamma Q_L=Q_L$,
$\bar \Gamma Q_R=Q_R$, with $\bar \Gamma=\Gamma_0\Gamma_1\cdots \Gamma_9$.

Consider first a Dirichlet fivebrane
located at $x^6=x^7=x^8=x^9=0$.  It is invariant under linear combinations
$\epsilon^LQ_L+\epsilon^RQ_R$ of supersymmetries with $\epsilon^L$, 
$\epsilon^R$
being spinors such that
\beq
\epsilon_L=\Gamma_0\Gamma_1\Gamma_2\Gamma_3\Gamma_4\Gamma_5\epsilon_R.
\label{dfive}
\eeq
Of course, the unbroken supersymmetry is the same if the fivebrane is located
at any constant values of $x^6\,\ldots,x^9$, not necessarily zero.

Now consider an NS fivebrane located at definite values of 
$x^6\,\ldots\,x^9$. 
 It is likewise
invariant under half of the supersymmetries, namely those with
\beq
\epsilon_L=\Gamma_0\Gamma_1\cdots \Gamma_5\epsilon_L,~~\epsilon_R=-\Gamma_0
\Gamma_1\cdots \Gamma_5\epsilon_R.
\label{nsfive}
\eeq
One way to obtain this result is to apply an $SL(2,\Z)$ duality transformation
which converts the Dirichlet fivebrane to an NS fivebrane and maps
(\ref{dfive}) to (\ref{nsfive}).
Another approach is to directly study the conformal field theory of the 
fivebrane,
which is known \cite{CHS,CHS:Worldsheet,CHS:Worldbrane} not exactly but well
enough to
identify the unbroken supersymmetries.  In this description it is obvious
that the conformal field theory has no mixing between left- and right-moving 
modes,
which is the reason that $\epsilon_L$ and $\epsilon_R$ obey independent 
conditions
in (\ref{nsfive}).  This together with Lorentz invariance determines the form 
of 
(\ref{nsfive})
up to signs, and with one choice of what we mean by an NS fivebrane (as 
opposed
to an anti-fivebrane) the signs are as in (\ref{nsfive}).

Now we want to consider a supersymmetric configuration with both NS and 
Dirichlet
fivebranes.  A comparison of the above formulas shows that if both types
of fivebrane are present at definite values of $x^6,\ldots,x^9$, then no 
supersymmetry
at all is preserved.  Suppose, however, that we have an NS fivebrane at 
definite values of $x^6,\ldots,x^9$ and a Dirichlet fivebrane at definite
values of $x^3,x^4,x^5,x^6$.  Then (\ref{nsfive}) is unchanged, but
(\ref{dfive}) becomes
\beq
\epsilon_L=\Gamma_0\Gamma_1\Gamma_2\Gamma_7\Gamma_8\Gamma_9\epsilon_R.
\label{newdfive}
\eeq
Comparing (\ref{nsfive}) and (\ref{newdfive}), we find that this sort of
configuration preserves one quarter of the supersymmetries.

There is one further kind of brane that can be introduced without any further
breaking of supersymmetry.  
If we combine (\ref{nsfive}) and (\ref{newdfive}), 
we
deduce that 
\beq
\epsilon_L = \Gamma_0\Gamma_1\Gamma_2\Gamma_6\epsilon_R.
\label{threebrane}
\eeq
Supersymmetries generated by parameters obeying this equation are precisely
those that are unbroken in the presence of a Dirichlet threebrane
whose world-volume spans the $x^0,x^1,x^2$, and $x^6$ directions and
is appropriately oriented (that is, it is a threebrane rather than
an antithreebrane, which would  give a minus sign in \ref{threebrane}).
So we can add such threebranes without any additional supersymmetry breaking.

The presence of all of these branes breaks the Lorentz group $SO(1,9)$ to
$SO(1,2)\times SO(3)\times SO(3)$, where $SO(1,2)$ acts on $x^0,x^1,x^2$,
one $SO(3)$ acts on  $\vec m=(x^3,x^4,x^5)$, 
and one $SO(3)$ acts on $\vec w=(x^7,x^8,x^9)$.  
We will call the $SO(3)$'s  respectively $SO(3)_V$ and $SO(3)_H$;
their double covers $SU(2)_V$ and $SU(2)_H$ 
will act, respectively,
as symmetries of the Coulomb and Higgs branches.  

{}From what we have said so 
far, we will be
considering NS fivebranes at definite values of $x^6$ and $\vec w$,
and D fivebranes at definite values of $x^6$ and $\vec m$.  We will
write $t_i$ and $\vec w_i$ for the values of $x^6$ and $\vec w$ of the
$i^{th}$ NS fivebrane, and $z_j$ and $\vec m_j$      for the values
of $x^6$ and $\vec m$ of the $j^{th}$ D fivebrane.

The threebranes that we consider will not be infinite threebranes;
we will consider threebranes ending on fivebranes.
(Starting with the fact that elementary strings can  end on
Type IIB  threebranes and applying various perturbative and
nonperturbative dualities, one can deduce \cite{Strominger:1996,Townsend:1996}
that threebranes can end on arbitrary fivebranes.)  Three kinds
of threebrane will appear: those with both ends  on
an NS fivebrane, those with both ends on a D fivebrane,
and those with one end on one kind of fivebrane and the other
on the other kind of fivebrane.  

To preserve supersymmetry, a threebrane must have      its world-volume
precisely in the $x^0,x^1,x^2,x^6$ directions, so a threebrane
can connect two given fivebranes in a supersymmetric configuration
only if the transverse positions of the fivebranes obey certain
constraints.  To be precise, two NS fivebranes $i$ and $i'$ can be
connected by a threebrane if and only if $\vec w_i=\vec w_{i'}$,
and a threebrane connecting them has an arbitrary value of $\vec m$
(since the NS fivebrane world-volumes range over all values of
$\vec m$). We will write $\vec x_\alpha$ for the value of $\vec m$
of the $\alpha^{th}$ threebrane.
Likewise, two D fivebranes $j$ and $j'$ can be connected by
a threebrane if and only if $\vec m_j=\vec m_{j'}$, and a threebrane
connecting them has an arbitrary value of $\vec w$.  We will write
$\vec y_\alpha$ for the value of $\vec w$ of the $\alpha^{th}$
threebrane.  Finally, any
NS fivebrane can be connected to any D fivebrane by a threebrane,
since the NS fivebrane spans all $\vec m$ and the D fivebrane spans
all $\vec w$.  But there are no moduli in the position of a
threebrane connecting an NS fivebrane to a D fivebrane; the $\vec w$
value of such a threebrane is that of the NS fivebrane, and
its $\vec m$ value is that of the D fivebrane.

A virtue of this setup, which will ultimately lead to our explanation
of mirror symmetry, is that there is a symmetry between the two
kinds of fivebrane.  In fact, an $SL(2,\Z)$ transformation by
the matrix 
\beq
S=\pmatrix{0 & 1 \cr -1 & 0 } 
\label{foofo}
\eeq
exchanges the two kinds of fivebrane.
Let R be 
 a rotation that maps $x^j$ to $x^{j+4}$ and 
$x^{j+4}$ to $-x^j$ for $j=3,4,5$ and leaves other coordinates
invariant.  The combined operation RS 
maps the class of configurations we are considering to itself,
while exchanging the two kinds of fivebrane.  We will call this
mirror symmetry, justifying the name in due course.

\section{Field theory on the D3 brane}\label{sec:matter}

Our general point of view in studying this problem is that because
the fivebranes are infinite in two directions not shared by the
threebranes, we can think of the fivebranes as being much heavier
than the threebranes.  Therefore, we think of the fivebrane
parameters as being fixed, and we study the quantum dynamics of the
threebrane motion, by analogy with many other investigations of
field theory on branes, such as \cite{BANKS,NS}.
The parameters specifying the fivebrane positions -- $z$, $t$, $\vec w$,
and $\vec m$ -- will be interpreted as coupling constants in
an effective quantum field theory on the threebrane world volume.
The positions of the D3 branes are dynamical
moduli which parametrize the vacua for the field theory. 

Since each threebrane we consider has only three infinite directions
$x^0,x^1$, and $x^2$ (and spans only a finite extent in $x^6$),
the threebrane field theory is macroscopically $2+1$-dimensional.
This is rather like Kaluza-Klein theory where one compactifies on a circle
and thereby reduces to a theory with a smaller number of macroscopic
spacetime dimensions.  The number of 
supercharges preserved by the configuration is $8$ 
so we are dealing with  $N=4$ supersymmetry in three dimensions.
$N=4$ supersymmetry in 
three dimensions has potentially an $SO(4)=SU(2)_V\times SU(2)_H$
R-symmetry.   For the class of models we are considering,
those symmetries are actually present: they can
be identified with the groups $SU(2)_V$ and $SU(2)_H$ already introduced
in the last section.  Incidentally, while there are many ways
to embed  a low energy field theory in string theory, in most
instances the $R$ symmetries -- which are very important in the
field theory dynamics -- are only approximate symmetries of the string
theory.  The fact that in the approach considered here the $R$
symmetries are visible directly in string theory is one reason
for the power of the construction.

To determine {\it which} $2+1$-dimensional field theories we obtain
by this construction, we begin by considering a single {\it infinite}
threebrane.  We will write $N$ for the number of supersymmetries
counted in three dimensions and ${\cal N}$ for the number counted
in four dimensions.  The world-volume theory on an infinite threebrane
is a four-dimensional theory with twice as much supersymmetry
as the models we will actually be studying in this paper;
it has ${\cal N}=4$ supersymmetry in the four-dimensional sense
or $N=8$ from a three-dimensional point of view.  The field
theory on the infinite threebrane is  a $U(1)$ gauge theory with
a supermultiplet that is irreducible under $N=8$ (or ${\cal N}=4$).
But under the $N=4$ subalgebra that will actually be a symmetry
of the models we will consider (in which the threebranes end on
fivebranes), the ireducible $N=8$ multiplet decomposes as the sum of a 
vector multiplet and a hypermultiplet.

When a threebrane ends on a fivebrane, 
there are boundary conditions that set to zero half of the
massless fields on the threebrane world-volume.  It is important
to know which half.  Supersymmetry alone would allow boundary
conditions in which either the vector multiplet or the hypermultiplet
vanishes on the boundary (that is, obeys Dirichlet boundary conditions),
while the other is free (obeys Neumann boundary conditions).
In more detail, the possible boundary conditions are as follows.

\begin{enumerate}
\item A scalar field in $3+1$ dimensions can obey either Dirichlet
boundary conditions, in which the scalar vanishes on the boundary,
or Neumann boundary conditions, in which the boundary values are
unconstrained but the normal derivative vanishes on the boundary.

\item A vector field in $3+1$ dimensions may obey either Dirichlet
boundary conditions, in which the components of $F_{\mu\nu}$ with
$\mu$ and $\nu$ tangent to the boundary vanish, or Neumann boundary
conditions, in which the components in which one index is tangent
to the boundary vanish.  Note that a vector field $A$ in $3+1$ dimensions
reduces to two fields in $2+1$ dimensions.  If (as in the above
conventions), we call the normal coordinate to the boundary $x^6$,
and the other coordinates $x^\mu$ ($\mu=0,1,2$), the two $2+1$-dimensional
fields are a scalar $b$ such that $\partial_\mu b=F_{\mu 6}$ and
a $2+1$-dimensional $U(1)$ gauge field $a$ such that $a_\mu$ for 
$\mu=0,1,2$  is the $x^6$-independent component of $A_\mu$. 
(In the free abelian theory, one can make a duality transformation
to turn  $a_\mu$ into another scalar.) 
Neumann boundary conditions on $A$ set the scalar $b$ to zero in the
effective $2+1$-dimensional theory, and Dirichlet boundary conditions
set the vector $a$ to zero.  Note that any transformation that
acts as electric-magnetic duality on $A$ exchanges the 
components of $F_{\mu\nu}$ with both indices tangent to the boundary
with components with one normal index, and so exchanges
$b$ with the scalar
dual to $a_\mu$ and  exchanges the two types of boundary
condition on $A$.

\end{enumerate}

The potential massless modes in the effective $2+1$ dimensional
theory on a threebrane are  the fluctuations in the transverse
position $\vec x$ and $\vec y$ (introduced in the last section)
plus the scalar $b$ and the vector $a_\mu$.  Under the $N=4$
algebra described in equations (\ref{dfive}) - (\ref{threebrane}), $\vec x$ and
$a_\mu$ form the bosonic part of a vector multiplet, while $\vec y$ and $b$
form the bosonic part of a hypermultiplet.  Therefore,  
when a threebrane ends on one of the fivebranes introduced
above, the boundary conditions are such that the bosonic
modes not set to zero
on the boundary are either  $\vec x$ and $a_\mu$ or $\vec y$ and $b$.

At this point it is easy to deduce what happens.  When a threebrane
ends on an NS fivebrane, $\vec x$ is not set to zero because
it is free to fluctuate in the case of a threebrane suspended
between {\it two} fivebranes, as described in the next to last paragraph
of the previous section.  So 
 for a threebrane ending on an NS fivebrane there are boundary
conditions which set to zero
$\vec y$ and $b$ and leave $\vec x$ and $a_\mu$ as massless modes.
Conversely, $\vec y$ is not set to zero when a threebrane ends on a
D fivebrane, as it is free to fluctuate for a threebrane suspended
between two D fivebranes.  So the boundary conditions
for a threebrane ending on a D fivebrane set to zero $\vec x$ and $a_\mu$
and leave $\vec y$ and $b$ as massless modes.

Note that these statements are compatible with the 
mirror symmetry transformation
introduced at the end of the last section.  This exchanges the 
two kinds of fivebrane, exchanges $\vec x$ with $\vec y$, and
(because it acts as electric-magnetic duality on the threebrane
theory \cite{Tseytlin:1996,Green:1996}) exchanges $b$ with the scalar
dual to $a_\mu$.

Now the effective $2+1$ dimensional theory on a threebrane
can be identified.

\begin{enumerate}
\item
If a threebrane has  both ends on a NS fivebrane, 
the effective $2+1$-dimensional theory is that of a $U(1)$ vector
multiplet.  More generally, given $n_v$ parallel threebranes suspended
between the same two NS fivebranes, we get $n_v$ vector multiplets;
the $U(1)^{n_v}$ gauge symmetry is at the ``classical level'' enhanced
to $U(n_v)$ via Chan-Paton factors when the parallel threebranes
become coincident.  What really happens in the regime in which they
are nearby is the problem of understanding the dynamics of the Coulomb
branch of the $U(n_v)$ quantum gauge theory.

\item
Similarly we consider the RS dual or mirror configuration. 
The world-volume theory for a threebrane stretched between two
D fivebranes is that of a massless hypermultiplet.
Given $n_h$ parallel threebranes suspended between the same 
two D fivebranes, we get $n_h$ massless hypermultiplets as long
as their positions are far enough apart; what actually happens when
they are close must be elucidated.
By an RS duality transformation turning the D fivebranes into
NS fivebranes, the hypermultiplets are converted into vectors
and the $n_h$ massless hypermultiplets parametrize the Coulomb
branch of a $U(n_h)$ gauge theory.

\item
The last case is that of a threebrane which terminates on
a D fivebrane at one end and an NS fivebrane at the other end.
Such a threebrane has no moduli. 
$\vec x$ is fixed at one end to equal the $\vec m$ value of the
D fivebrane (and
its supersymmetric partner $a_\mu$ is projected out by Dirichlet
boundary  conditions at that end) and
$\vec y$ is fixed (along with $b$)
at the other end.  For a single such threebrane, there
are no massless modes on the worldvolume at all: the low energy
theory on a single infinite threebrane is an infrared-free $U(1)$
gauge theory, and all massless modes are projected out at one end or the
other.  The infrared theory has a unique vacuum with a mass gap.
\end {enumerate}

The gauge group arising on a threebrane stretched between two
NS fivebranes will be called an electric gauge group in what follows.
The gauge coupling of the electric gauge theory is easy to identify.
If the $x^6$ values of the two NS fivebranes are $t_1$ and $t_2$
then up to a universal multiplicative constant that depends only
on the coupling constant $\tau_{IIB}$ 
of the underlying Type IIB superstring theory,
the coupling of the effective $2+1$-dimensional field theory is
\beq
{1\over g^2}=|t_1-t_2|.
\label{fluff}
\eeq
In fact, the four-dimensional effective coupling $g_4$ on a threebrane
is determined universally in terms of $\tau_{IIB}$.  The three-dimensional
effective coupling, obtained by integrating over $x^6$ to reduce to a
three-dimensional effective Lagrangian, is 
\beq
{1\over g^2}={|t_1-t_2|\over g_4^2},
\label{uff}
\eeq
and this is the basis for (\ref{fluff}).

Likewise, if we perform an RS duality transformation to interpret
the fields on a threebrane that ends on two D fivebranes as vector
multiplets, we get a gauge theory with a gauge group that we will
call a magnetic gauge group.  If the $x^6$ values of the D fivebranes
are $z_1$ and $z_2$, then by reasoning just as above, the magnetic
gauge coupling is
\beq
{1\over g^2} =|z_1-z_2|.
\label{nuff}
\eeq
In both (\ref{fluff}) and (\ref{nuff}) we omitted an overall constant; those
constants
are in fact equal if the Type IIB coupling is equal to the duality-invariant
value $\tau_{IIB}=i$.

\begin{figure}[htb]
\centerline{\psfig{figure=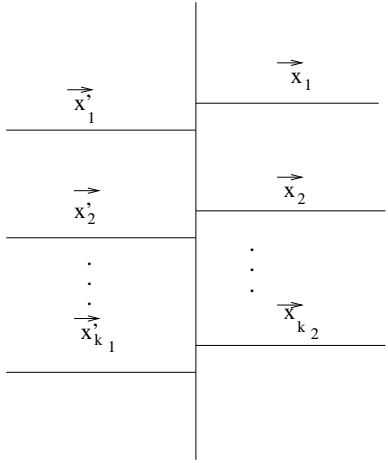,width=2in}}
\caption[]{
Here and in subsequent figures,
 vertical solid lines represent NS  fivebranes in the $012345$ directions,
and horizontal lines represent $0126$ threebranes.  In the example
depicted here, the threebranes come from left or right, and a massless
hypermultiplet appears whenever a ``left'' and ``right'' threebrane meet.
\protect\label{fig:NS}}
\end{figure}

\def\bar{\overline}
We now want to discuss certain singularities that can result in the
appearance of additional massless hypermultiplets in the effective
$2+1$-dimensional theory.  First consider a
 solitonic or NS fivebrane such that $k_1$ threebranes end on it from the
left and $k_2$ threebranes  end on it from the right, as in figure \ref{fig:NS}.

We want to consider what happens when one of the ``left'' threebranes
meets one of the ``right'' threebranes, that is, when their $\vec x$
values coincide and they actually meet in space-time.  
This is a sort of singularity
at which extra massless states might appear.  In this case
it is easy to guess heuristically what might happen.  
A string stretched
between the ``left'' threebrane and the ``right'' threebrane 
could give a hypermultiplet that becomes
massless in the limit that the two threebranes meet.
To be somewhat more precise, if the $\vec x$'s of the two threebranes
differ by a vector of length $\alpha$ that points in the $x^3$ direction
 then an elementary string stretched
between the two threebranes would appear to give a BPS hypermultiplet
of mass $\alpha$, 
invariant under supersymmetries generated by parameters
that obey
\beq
\epsilon_L=\Gamma_0\Gamma_3
\epsilon_L, \qquad\epsilon_R=\Gamma_0\Gamma_3\epsilon_R,
\eeq
as well
as equations (\ref{dfive}) - (\ref{threebrane}).  
This hypermultiplet will
become massless when the two threebranes meet.
The existence of this state is not really a sound deduction from
perturbative string theory -- as the configuration of figure \ref{fig:NS}
is beyond the reach of perturbative string theory -- and we will
eventually give additional arguments that this state must be present.

The  hypermultiplet obtained in this way  contains four real
scalars which transform in the $(1,2)$ representation 
of the R-symmetry group
$SU(2)_V\times SU(2)_H$ (the representation $(1,2)$ is four-dimensional
if viewed as a real representation) plus fermions.
The coupling of this  hypermultiplet to the ``electric'' vector
multiplets on the threebranes gives it 
a mass proportional to  $|\vec x_L-\vec x_R|$ where $\vec x_L$ and
$\vec x_R$ are the $\vec x$ values of the ``left'' and ``right'' threebrane.

Now consider the full configuration of figure \ref{fig:NS} with
many $k_1$ ``left'' threebranes and $k_2 $ ``right'' ones.
As there are $k_1k_2$ possible coincidences, involving the
meeting of a ``left'' threebrane with a ``right'' one, there are
 a total of $k_1k_2$
hypermultiplets associated with these singularities.
 In view of their couplings to the $U(1)$ gauge fields on the various
threebranes, they clearly
transform as $(k_1,\bar k_2)$ under $U(k_1)\times U(k_2)$.  

\begin{figure}[htb]
\centerline{\psfig{figure=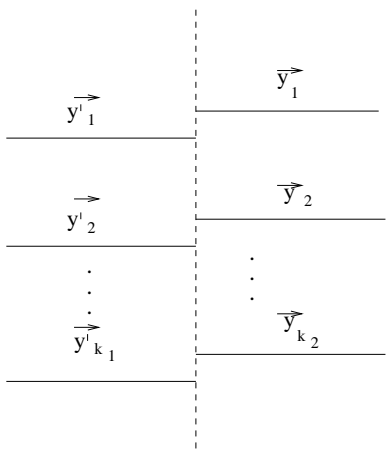,width=2in}}
\caption[]{Here and in subsequent figures,
vertical dashed lines represent D fivebranes 
in the $012789$ directions (which one might think of as coming
``out of the paper''),  while horizontal lines
represent D3 branes which in this example come from the left or right.
\protect\label{fig:D5}}
\end{figure}

Next let us consider the mirror configuration, in which we will
get hypermultiplets that are charged with respect to the {\it magnetic}
gauge group.
Performing an RS  duality transformation which converts NS fivebranes
to D fivebranes while leaving the threebranes invariant,
the picture considered above is converted to that of  figure \ref{fig:D5}.
In this case, coincidences of ``left'' and ``right'' threebranes give massless
hypermultiplets transforming as $(k_1,\bar k_2)$ of the {\it magnetic}
gauge group.  Under the $SU(2)_V\times SU(2)_H$ symmetry, the
scalars in these hypermultiplets transform as $(2,1)$.

Notice that these two mirror-symmetric constructions gives hypermultiplets
that transform differently under $SU(2)_V\times SU(2)_H$, in fact
as $(1,2)$ and $(2,1)$ respectively.  When it is necessary to draw
a distinction, we will call the fields obtained in this way (and charged
with respect to electric or magnetic gauge groups) 
electric hypermultiplets and magnetic hypermultiplets, respectively.

\begin{figure}[htb]
\centerline{\psfig{figure=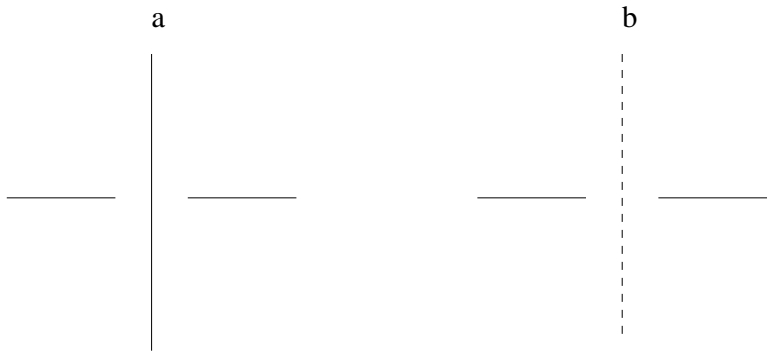,width=4in}}
\caption[]{Electric and magnetic hypermultiplets. Sketched
in  figure (a)  is a threebrane (horizontal line) perpendicular
to an NS fivebrane (vertical line).  When they actually meet
in space (that is, when
 the $\vec m$ values are equal), a massless hypermultiplet
appears.  Sketched 
in figure (b) are
a D fivebrane (vertical dotted line) and perpendicular  threebrane 
(horizontal line); a massless magnetic
hypermultiplet appears when they actually meet in space.  In later
figures, we will not be so careful in indicating whether branes
represented by crossing perpendicular lines actually meet in space.
\protect\label{fig:NS5D3}}
\end{figure}

There is another situation in which hypermultiplets appear.
Consider as in figure \ref{fig:NS5D3}
a threebrane perpendicular to a D fivebrane or an NS fivebrane.  
A massless hypermultiplet
appears when the threebrane and fivebrane actually meet in spacetime,
which occurs when the threebrane transverse position is suitably adjusted.
In figure \ref{fig:NS5D3}(b), where the fivebrane is Dirichlet, this can be
demonstrated explicitly in weakly coupled string theory by considering an
elementary Type IIB superstring suspended between the threebrane and the
D fivebrane. This gives a  hypermultiplet with scalar fields
transforming as $(1,2)$ 
under $SU(2)_V\times SU(2)_H$ and mass parameters equal
to $\vec x-\vec m$.  By 
mirror symmetry, the configuration of figure \ref{fig:NS5D3}(a), with a 
solitonic
or NS fivebrane, therefore gives a hypermultiplet with
scalars transforming as $(2,1)$ under $SU(2)_V\times SU(2)_H$ and
mass parameters $\vec y-\vec w$.

Notice that we have now given two different ways of obtaining such light
``electric'' hypermultiplets, involving the configurations of figures
\ref{fig:D5} and \ref{fig:NS5D3}(b).  In the second, the appearance of the
light hypermultiplet
is a precise deduction of perturbative string theory, and in the first
it is not.  Later, we will understand some ``phase transitions'' in which
the second mechanism
of generating a light electric hypermultiplet is converted into the first,
giving a powerful check that the situation of figure \ref{fig:NS} behaves
as claimed.  One of these phase transitions is based on an esoteric
process involving motion of fivebranes, but one can be described
very simply.   Consider as in figure 3 a threebrane perpendicular
to a fivebrane, and adjust the threebrane position so that it meets
the fivebrane in space; it is certainly plausible intuitively that
the threebrane can then ``break'' into two pieces, one meeting the
fivebrane from the left and one from the right, as in figure 1.  If
such a  transition between the configurations of figures 1 and     3 is
possible -- and in section five we will see that such a transition is
needed to reproduce certain standard field theory phenomena -- then,
since figure 3 generates a massless hypermultiplet at the point at which
the different branes meet in spacetime, figure 1 must do the same.

\begin{figure}[htb]
\centerline{\psfig{figure=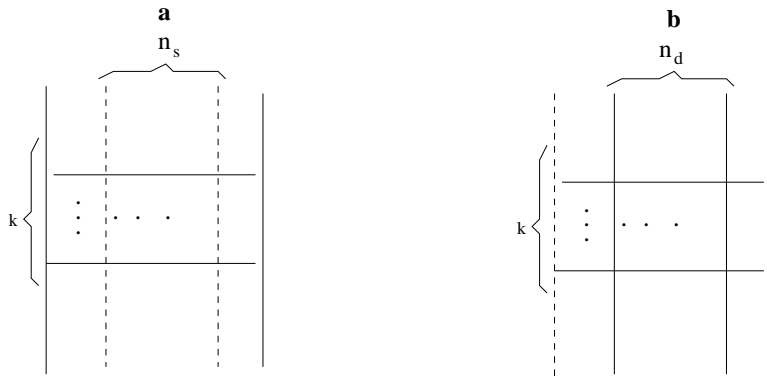,width=4in}}
\caption[]{Hypermultiplets in electric and magnetic gauge theories.
In part (a),  electric hypers are given by threebrane
intersections with  D fivebranes. In part (b), magnetic
hypers come from intersections NS fivebranes.
\protect\label{fig:twoNSnD5kD3}}
\end{figure}

A particularly important application of the construction described in the
last two paragraphs is as follows.
Consider  as in figure \ref{fig:twoNSnD5kD3}(a)
two solitonic five branes with $x^6$ positions $t_1$ and $t_2$, with
$t_1<t_2$,
and with $k$ threebranes  stretching in
between them.   Such a supersymmetric
configuration exists if and only if the two solitonic fivebranes have
 $\vec w_1=\vec w_2$.  Suppose that in addition there is present
as well a D fivebrane with $x^6$ position $t_1<z<t_2$.
Then by what we have just said,
a massless hypermultiplet appears whenever a threebrane meets
a D fivebrane, that is whenever the $\vec x$ value of the threebrane
equals the $\vec m$ transverse position of the D fivebrane.

If there are $k$ parallel threebranes in figure \ref{fig:twoNSnD5kD3}(a), then
the effective world-volume theory is a $U(k)$ gauge theory, and the
intersections with the D fivebrane will give $k$ potentially massless
hypermultiplets transforming in the fundamental representation of $U(k)$.
Such a multiplet can have a supersymmetric bare mass, transforming 
as $(3,1)$ of $SU(2)_V\times SU(2)_H$.   In this case, the bare mass
is equal to the transverse location $\vec m$ of the Dirichlet fivebrane.
For a single hypermultiplet in the fundamental representation of $U(k)$,
this bare mass is not very exciting, as it can be canceled by moving
the $\vec x$ values of the threebranes, that is, by shifting the
scalars in the $U(k)$ vector multiplet.  

Suppose more generally that as in the figure there are $n_d$
D fivebranes whose positions are $z_i,\vec m_i$, $1\leq i\leq n_d$,
with $t_1<z_i<t_2$.  Intersections
with the threebranes now give hypermultiplets transforming as $n_d$ copies
of the fundamental representation of $U(k)$.  The bare mass of the
$j^{th}$ such multiplet is $\vec m_j$, and the differences $\vec m_i-\vec m_j$
are observable parameters of the low energy effective theory in $2+1$ 
dimensions;
they cannot be  eliminated by shifting any of the dynamical fields.

Since the picture of figure \ref{fig:twoNSnD5kD3}(a) corresponds to a $U(k)$ 
gauge
theory,
and $U(k)$ has a non-trivial center $U(1)$, a Fayet-Iliopoulos $D$-term
for the $U(1)$  should be possible.  This is simply the difference
$\vec D=\vec w_1-\vec w_2$ between the transverse positions of the two
NS fivebranes at the ends of the figure.  We already noted that a 
supersymmetric
configuration of the type sketched in figure \ref{fig:twoNSnD5kD3}(a) only 
exists if
$\vec w_1-\vec w_2=0$.  For $\vec D\not= 0$, a supersymmetric vacuum may still
exist, but only after making a transition to a Higgs branch such as we
discuss later.  

Finally, we consider the 
mirror of this.  This is obtained, of course, by converting all NS
fivebranes into D fivebranes, and vice-versa, to arrive at
figure \ref{fig:twoNSnD5kD3}(b).
Now we get massless hypermultiplets transforming in the fundamental
representation of a magnetic $U(k)$ gauge group; the massless scalars
transform as $(2,1)$ of $SU(2)_V\times SU(2)_H$, and the bare mass parameters
are the $\vec w$ parameters of the NS fivebranes.  The difference
$\vec m_1-\vec m_2$ for the D fivebranes at the ends of the figure
is the Fayet-Iliopoulos $D$-term for the magnetic $U(k)$ gauge theory.

\section{The Coulomb Branch And The Moduli Space Of Monopoles}

We finally move on to applications.
In this section, we  consider the relatively tame situation in which all
gauge groups are electric.  As we will see, the phenomena are still
quite rich and interesting.

\begin{figure}[htb]
\centerline{\psfig{figure=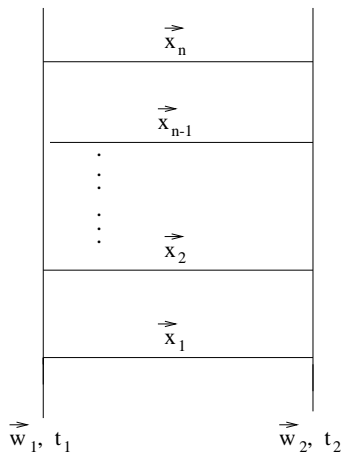,width=2in}}
\caption[]{$SU(2)$ monopoles versus pure $N=4$ $U(k)$ three dimensional gauge
theories. The vertical lines represent solitonic five branes at the positions
$\vec w_1,t_1$ and $\vec w_2,t_2$. The horizontal lines represent D3 branes
with positions $\vec x_i, i=1,\ldots,k$.
\protect\label{fig:YM}}
\end{figure}

First we consider, as in figure \ref{fig:YM},
a system of $k$ parallel threebranes suspended between a pair of
NS fivebranes.  To an observer on the threebranes, the low energy
theory is, as we have seen in the last section, a $U(k)$ electric
gauge theory with no hypermultiplets.  The transverse positions of the 
threebranes (together with their superpartners,
 the scalars dual to the vector fields on the threebranes) parametrize
the Coulomb branch of this theory.

On the other hand, the same configuration can be viewed in another way.
A $5+1$-dimensional observer sees a $U(2)$ gauge theory of two
parallel fivebranes, with $U(2)$ broken to $U(1)\times U(1)$ by the separation
between these branes.  Actually the center  $U(1)\subset U(2)$ will play no 
role in this discussion -- as the light fields are all neutral under
this $U(1)$ -- so we can work modulo the center and think of the
fivebranes as carrying a $SU(2)$ gauge symmetry that is broken to $U(1)$.
The Higgs field that controls this symmetry breaking is simply
$t_1-t_2$, the separation between the two fivebranes in the $x^6$ coordinate.
The fivebrane theory of $SU(2)$ broken to $U(1)$ is to be treated
classically as the fivebranes are so heavy.

Now from the point of view of the fivebrane theory, the end of a threebrane
looks like a magnetic monopole.  (To be more precise, this might be called
a magnetically charged twobrane; a magnetic monopole is a particle in
$3+1$ dimensions or a twobrane in $5+1$ dimensions.  However, we will
be treating the fivebrane theory classically and our 
configurations
will all be invariant under translations of $x^1$, and $x^2$, so it
is reasonable to call them magnetic monopoles in a $3+1$-dimensional
reduction of the fivebrane world-volume in which $x^1$ and $x^2$
are suppressed.)  Since in figure \ref{fig:YM}
 we have $k$ threebranes ending on the fivebranes, the configuration
looks like one of magnetic charge $k$ from the point of view of the fivebrane
observer.

Because of the unbroken supersymmetry of the brane configuration
in figure \ref{fig:YM}, the fivebrane observer sees, to be more precise, a
BPS-saturated configuration with magnetic charge $k$ which depends
on a total of $4k $ real moduli -- the same parameters (threebrane
transverse positions and scalars dual to threebrane world-volume vectors)
which we earlier interpreted as parametrizing the Coulomb branch of the
$U(k)$ gauge theory.  These $4k$ variables parametrize a hyper-Kahler
manifold ${\cal M}$.

Thus, we have something that may sound familiar: a  gauge theory with
$SU(2)$ broken to $U(1)$, and a $4k$-dimensional hyper-Kahler manifold
parametrizing BPS-saturated configurations of magnetic charge $k$.
The classical moduli space ${\cal M}_{cl}$ of BPS monopoles of magnetic
charge $k$ has exactly these properties \cite{Atiyah:1988}, and it
is natural to suspect that ${\cal M}={\cal M}_{cl}$.  In fact, essentially
this question has been analyzed by Diaconescu \cite{Dia}.  After
making a mirror transformation to convert the NS  fivebranes of
figure \ref{fig:YM}
to D fivebranes, followed by a $T$-duality transformations to reduce to
onebranes ending on threebranes, 
the configuration of figure \ref{fig:YM} turns into the system of strings
ending on threebranes
considered by Diaconescu, who analyzed the string dynamics and obtained
the Nahm equations for monopoles, thus showing that 
${\cal M}={\cal M}_{cl}$.

So at this point, we have learned that one moduli space -- that of
figure \ref{fig:YM} -- can be looked at in two ways.  It is the Coulomb branch
of the pure $U(k)$ quantum 
gauge theory in three dimensions, without hypermultiplets,
or it is the moduli space of $k$ monopoles in a classical gauge theory 
in three dimensions with $SU(2)$ broken down to $U(1)$.
This correspondence has been previously noted in \cite{SW} for $k=2$ and
in \cite{CH} for general $k$.

Notice that, as the center $U(1)\subset U(k)$ decouples from the $U(k)$
gauge theory, the moduli space considered here is a product of the
Coulomb branch of an $SU(k)$ gauge theory with the moduli space of
a free vector multiplet.
 From the other point of view, the free vector multiplet parametrizes the
center of mass of the $k$-monopole configuration.   An alternative statement
is thus that the reduced moduli space of the $k$-monopole system,
with the center of mass position factored out, is the Coulomb branch
of the supersymmetric $SU(k)$ gauge theory with no hypermultiplets.

\subsection{The General Case}

Now we will consider a much more general situation of the same kind.

We consider a $U(n)$ gauge theory with $U(n)$ broken to $U(1)^n$
by the expectation value of a Higgs field $\phi$ which takes values
in the adjoint representation; that is, $\phi$ is an $n\times n$ Hermitian
matrix.  The symmetry breaking is parametrized by the eigenvalues of
$\phi$ which we call $\phi_i$, with $\phi_1< \phi_2< \cdots \phi_n$.
(We could replace $U(n)$ by $SU(n)$ and add a constant to $\phi$ so
that $\sum_i\phi_i=0$; this would bring no essential change in what follows.)

A basic $SU(2)$ BPS monopole can be embedded in $U(n)$ in $n-1$ different
ways, to give monopoles of magnetic charge $(1,-1,0,\ldots, 0)$,
$(0,1,-1,\ldots, 0), \ldots$, or $(0,0,\ldots,1,-1)$.    
By combining $k_i$ monopoles of the $i^{th}$ type for $1\leq i\leq n-1$
we build monopoles of charges $(k_1,k_2-k_1,\dots, -k_n)$ and the
general BPS configuration 
is parametrized by the choice of the non-negative
integers $k_i$ as well as of the Higgs eigenvalues $\phi_i$
\cite{Weinberg:1980,Murray}.

\begin{figure}[htb]
\centerline{\psfig{figure=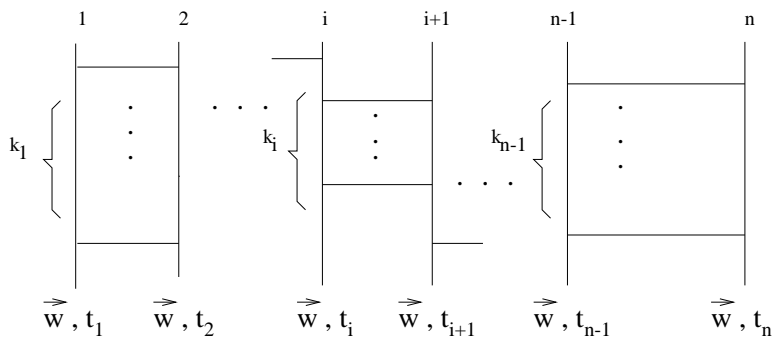,width=4in}}
\caption[]{$SU(n)$ monopoles versus three dimensional gauge theories.
\protect\label{fig:SUnmon}}
\end{figure}

Now we can realize this 
general component of $U(n)$ monopole moduli space via a
brane configuration, as follows.  We consider, as in figure \ref{fig:SUnmon},
$n$ parallel NS fivebranes, with $x^6$ values $t_1,\dots,t_n$,
and with the $i^{th}$ fivebrane connected to the $i+1^{th}$ by
$k_i$ threebranes.  Such a configuration is possible precisely if
the fivebranes have a common value of $\vec w$, so the parameters
that enter in defining it are the continuous variables $t_j$ and the
integers $k_j$.

Looked at from the fivebrane point of view, the configuration of
figure \ref{fig:SUnmon} is a $U(n)$ monopole, with Higgs eigenvalues
$\phi_j=t_j$, and magnetic
charges given by the $k_j$.  The moduli space is thus the BPS moduli
space of such objects, which we will call ${\cal M}(k_j;\phi_j)$.
On the other hand, from the threebrane point of view, we have macroscopically
a $2+1$ dimensional supersymmetric gauge theory with gauge group
$U(k_1)\times U(k_2)\times \cdots \times U(k_{n-1})$ and with, in an obvious
notation, 
hypermultiplets of zero bare mass transforming in the representation
${\bf (k_1,\bar k_2)\oplus (k_2,\bar k_3)\oplus \cdots \oplus(k_{n-2},\bar 
k_{n-1})}$.
In this gauge theory, the gauge coupling of the $j^{th}$ factor $U(k_j)$
of the gauge group is given by
\beq
{1\over g_j^2}=|\phi_j-\phi_{j+1}|.
\label{gc}
\eeq
The Coulomb branch of this gauge theory must thus coincide, as a hyper-Kahler
manifold, with the BPS moduli space ${\cal M}(k_j;\phi_j)$.  Thus,
we have identified {\it all}  BPS moduli spaces for $U(n)$ gauge group
with the Coulomb branches of suitable gauge theories.

As a special case of this, suppose that there are precisely
three fivebranes, so that the gauge group is $U(k_1)\times U(k_2)$
and the hypermultiplets transform as ${\bf (k_1,\bar k_2)}$.
The gauge couplings of $U(k_1)$ and $U(k_2)$ are respectively
\beq
{1\over g_1^2}= t_2-t_1=\phi_2-\phi_1
\label{onec}
\eeq
and
\beq
{1\over g_2^2} =t_3-t_2=\phi_3-\phi_2.
\label{oned}
\eeq
The two gauge couplings are completely independent, and if
we wish we can take $g_2\to 0$, keeping $g_1$ fixed, by taking
$\phi_3\to\infty$, with $\phi_2-\phi_1$ fixed.
In the limit that $g_2$ is turned off, the theory, from the
threebrane point of view, turns into a  $U(k_1)$ gauge theory
with  hypermultiplets transforming as $k_2$ copies of 
the fundamental representation of 
$U(k_1)$; the $U(k_2)$ gauge symmetry
reduces to a  global $U(k_2)$ symmetry acting on $k_2$ multiplets.
The hypermultiplets have arbitrary bare masses determined by the
expectation values of the scalars in the $U(k_2)$ vector multiplet.

Now, let us look at it from the fivebrane point of view.
Here we are dealing with $SU(3)$ monopoles.  There are two
basic $SU(3)$ monopoles, of magnetic charge $(1,-1,0)$ and $(0,1,-1)$.
The sizes of the two monopoles are proportional to $1/|\phi_2-\phi_1|$
and $1/|\phi_3-\phi_2|$, respectively.   Thus when
$\phi_3$ becomes very large with fixed $\phi_2-\phi_1$, the $(0,1,-1)$
shrinks while the $(1,-1,0)$ monopole retains a fixed size.
In our problem, we have $k_1$ of the $(1,-1,0)$ monopoles
combined with $k_2$ of the $(0,1,-1)$ monopoles.  In the limit
that $\phi_3\to\infty$ with $\phi_2-\phi_1$ fixed, one would
expect the $(0,1,-1)$ monopoles to shrink to point singularities,
so we should be left with an $SU(2)$ monopole solution of magnetic
charge $k_1$ with $k_2$ point singularities.  The moduli space
of such singular $SU(2)$ monopoles should coincide with the Coulomb
branch of the $U(k_1)$ gauge theory with hypermultiplets
transforming as $k_2$ copies of the fundamental representation. The
positions of the singularities correspond to the bare masses of the
hypermultiplets in the field theory. 

Kronheimer 
\cite{kronheimer} constructed a theory of $SU(2)$ monopoles
with point singularities and showed a close relation of
 the data involved to the data parametrizing $SU(3)$ monopoles.
Very plausibly, the singular monopoles considered by Kronheimer are
the ones relevant here.

\bigskip\noindent
{\it Explicit Comparison To Field Theory}

It should be possible to test the above  claims explicitly
by comparing computations of the metric on monopole moduli space
in the region in which the monopoles are widely separated to
computations of instanton corrections to the Coulomb branch of the
field theory.  The instantons in question are actually
monopoles (of the threebrane quantum gauge theory!), 
and can be reinterpreted in string theory in the following somewhat
subtle way.

\begin{figure}[htb]
\centerline{\psfig{figure=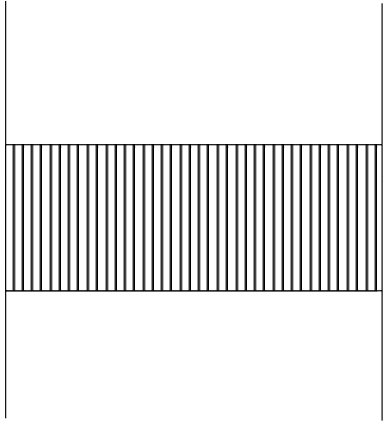,width=2in}}
\caption[]{Instanton corrections to the metric on the Coulmb branch
in $SU(2)$ gauge theory. These corrections
come from strings stretched between the two D3 branes.  The shaded
region is two-dimensional and is localized at specific values of
$x^0,x^1,$ and $x^2$ (the instanton position).
\protect\label{fig:instanton}}
\end{figure}

As a  simple relevant example, consider
the case that the fivebrane description involve $SU(2)$ monopoles
of magnetic charge two, and 
the  three-dimensional quantum gauge theory is an  $SU(2)$ gauge theory 
without hypermultiplets.
The corresponding string theory configuration consists of two parallel NS
fivebranes and two threebranes  stretching between them.
The two-monopole moduli space of $SU(2)$  is the Atiyah-Hitchin manifold,
studied in detail in \cite{Atiyah:1988}; its metric differs from a Taub-NUT
metric by terms that vanish exponentially at infinity.
We wish to
interpret the exponentially small corrections.  
According to \cite{SW}, these corrections are due to monopoles,
regarded as instantons of the $2+1$-dimensional theory, and
make contributions proportional to
$e^{-(I+i\sigma)}$, where $I$ is the action of the instanton and $\sigma$ is
the scalar dual to the $U(1)$ photon.
The action $I$ is proportional to ${|\vec x|\over g^2}$ where $\vec x$ are the
scalars in the vector multiplet while $g$ is the three dimensional coupling.
We need to look for a string theory instanton which give this action.
This action is nothing but the area spanned between the two D3
branes as in figure \ref{fig:instanton}.  

We thus claim a slight
enrichment of the general picture of branes ending on branes.
Not only can a D onebrane worldsheet end on a threebrane or a fivebrane;
it can also have a ``corner'' where the threebrane ends on a fivebrane.
Such onebrane worldvolumes with boundaries and corners are needed
to reproduce in string theory the field theory instanton corrections
to the metric on the two-monopole moduli space.

\subsection{A Special Example}

Finally, we consider in more detail another special example.

If all  $k_i$ equal
one, then the threebrane gauge theory is abelian, with
gauge group $U(1)^{n-1}$.    The
theory contains $n-2$ charged hypermultiplets, of charges
$(1,-1,0,\ldots, 0 )$, $(0,1,-1,0,\ldots,0), \cdots,
(0,0,\dots,0,1,-1)$.   Notice that the sum of the charges is zero for each
hypermultiplet, so after factoring out from the gauge group
a decoupled $U(1)$, this is a $U(1)^{n-2}$ gauge theory with
$n-2$ charged hypermultiplets.

For instance, the first case $n=3$ is the $U(1)$ theory with
one charged hypermultiplet, which was shown in \cite{SW} to have
for its Coulomb branch a smooth Taub-NUT manifold.  This indeed
agrees, as it should, with the $SU(3)$ monopole space for monopoles
of charge $(1,0,-1)$. This manifold was first studied in \cite{Connell} and
later discussed in \cite{Gauntlett:1996,Lee:1996}.

The fact that the gauge group is abelian means that there are no
instantons in the $2+1$-dimensional field theory.  Consequently,
the metric of the models    with all $k_i=1$
should be given exactly by a one-loop formula, with no
exponentially small corrections, and should be invariant under
shifts of the scalars dual to the photons.\footnote{These facts
are actually related in the following way.  The invariance
of the metric under shifts in the scalars actually means 
that the Coulomb branch admits what is called a tri-holomorphic
torus action of dimension equal to the number of vector multiplets.
This then implies \cite{Lindstrom:1983,Pedersen:1988} that the metric can
be written in a relatively elementary form using solutions of
linear equations.}
  
The Taub-NUT metric has these properties.

More generally, for arbitrary $n$,
the Coulomb branch of this theory should coincide
with the moduli space of $SU(n)$ monopoles of magnetic charge
$(1,0,\ldots,0,-1)$.  
The features reflecting absence of instanton corrections
have indeed been found in recent studies.
The metric on the moduli space of $(1,0,0,\ldots,-1)$ monopoles
  was conjectured in
\cite{Weinberg} and later proved by \cite{Murray,Gordon} to be
given by the formulas
\beq
ds^2=g_{ij}d\vec x_i\cdot d\vec x_j+(g^{-1})_{ij}d\tilde\theta_id 
 \tilde\theta_j,
\label{un}
\eeq
\beq
d\tilde\theta_i=d\theta_i+ 
  \sum_{j=1}^{n}\vec W_{ij}\cdot d\vec x_j . 
\label{thetan}
\eeq 
Here the function ${\vec W}_{ij}$ is the potential arising from a 
Dirac point 
monopole at the point $i$ acting at a monopole at the point $j$.
\beq 
W_{ii}=\sum_{i\not=j}w_{ij}, 
\qquad 
W_{ij}=-w_{ij},\qquad i\not=j.
\label{asymmet}
\eeq
\beq 
g_{ii}={|t_i-t_{i+1}|}+{g^2\over4\pi}\sum_{i\not=j}{1\over |\vec x_i-\vec x_j|}
\qquad 
g_{ij}=-{1\over |\vec x_i-\vec x_j|},\qquad i\not=j . 
\label{nondiag}
\eeq
This formula can be seen by symmetry arguments to agree with the Coulomb
branch of the $U(1)^{n-1}$ gauge theory with the one loop correction.
The decoupled $U(1)$ factor in $U(1)^{n-1}$ corresponds as usual
to 
the overall translational degree of freedom of the monopole system.  The
metric just given
can be interpreted as obtained 
from a flat metric by a one-loop correction; the
extra symmetries that arise because the threebrane theory 
has no instantons are simply the constant shifts in the $\theta$'s.

Incidentally, the bound state (or ${\bf L}^2$ harmonic form) on
this particular $SU(n)$ monopole moduli space 
that is predicted by duality and whose existence has recently been
confirmed, for the case $n=3$ by
\cite{Gauntlett:1996,Lee:1996} and for general $n$
by \cite{Gibbons:1996}, plausibly corresponds
in the threebrane language to a situation in which a single threebrane
stretches all the way between the leftmost and rightmost fivebranes,
without being split up into pieces.

\section {A First Look At Nontrivial Phase Trasitions}

Our aim is to incorporate both NS and D fivebranes and study the 
mirror symmetry that exchanges them.  First, though, we must
explain a new aspect of brane interactions that is needed
in order to study processes in which both kinds of fivebrane are present.

\begin{figure}[htb]
\centerline{\psfig{figure=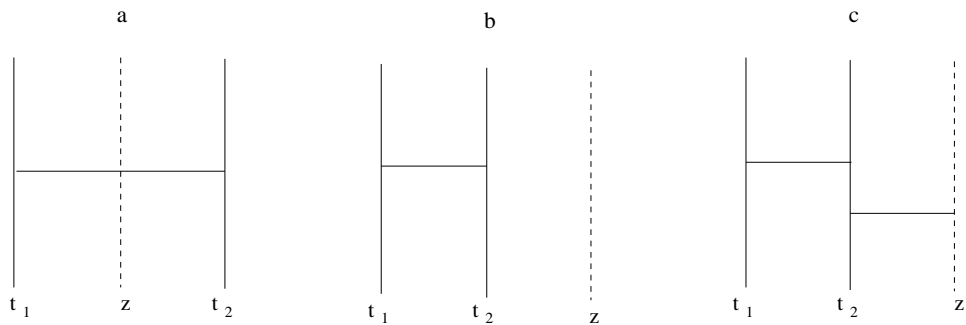,width=5in,angle=-90}}
\caption[]{Two fivebranes cross each other. In figure (a) the starting
configuration. In figure (b) the naive configuration obtained
by moving the D fivebrane of (a) to the right; consideration
of this transition leads to a paradox.
In figure (c) the correct configuration.
\protect\label{fig:paradox}}
\end{figure}

To see the need for a new phenomenon, we first describe an apparent
contradiction.  In figure \ref{fig:paradox}  we sketch two NS fivebranes
at $t_1,\vec w_1$ and $t_2,\vec w_2$, and a D fivebrane at $z,\vec m$.
There also is a threebrane connecting the two NS fivebranes, whose
transverse position      in the $3-4-5$ directions is $\vec x$.

First we consider the case that $t_1<z<t_2$, as in figure \ref{fig:paradox}(a).
In this figure there is a BPS hypermultiplet -- from strings connecting
the threebrane to the D fivebrane -- with a mass proportional to 
$|\vec m - \vec x|$.

Now gradually increase $z$ to get to the situation $t_1<t_2<z$.  Superficially,
the system should then be as depicted in figure \ref{fig:paradox}(b).  
Here, though,
there is a paradox: in figure \ref{fig:paradox}(b) there is no reason to get any
massless hypermultiplet (or any other singularity) at $\vec m-\vec x=0$.
In general, in models with three-dimensional $N=4$ (or four-diemnsional
${\cal N}=2$) supersymmetry, the BPS spectrum can jump, but in the
specific case at hand, there is nothing for the hypermultiplet
in question to decay to.

The resolution that we propose for this seeming paradox is as follows.
Note that at $z=t_2$, the D and NS fivebranes actually meet in spacetime.
In fact,
the D fivebrane is parametrized by the values of $x^0,x^1,x^2,x^7,x^8,x^9$,
with constant values of the other coordinates.  The NS fivebrane is
parametrized by the values of $x^0,x^1,x^2,x^3,x^4,x^5$ with constant
values of the others.  For them to meet in spacetime it is therefore
necessary and sufficient that the $x^6$ values coincide, and this
in particular occurs at $z=t_2$.  

Thus, in trying to deform figure \ref{fig:paradox}(a) to
figure \ref{fig:paradox}(b), the two
fivebranes have ``passed through'' each other.  Our proposal
is that when the two branes pass through each other, a third
brane is created.  In fact, we claim that
figure \ref{fig:paradox}(a) deforms not
to figure \ref{fig:paradox}(b) but to figure \ref{fig:paradox}(c).
In figure \ref{fig:paradox}(c), in addition
to the threebrane that was present in \ref{fig:paradox}(a),
there is a new threebrane,
running from the NS fivebrane at $x^6=t_2$ to the D fivebrane at $x^6=z$.
This new threebrane, since it connects an NS fivebrane to a D fivebrane,
has no moduli; its $\vec x$ is the $\vec m$ of the D fivebrane,
and its $\vec y$ is the $\vec w$ of the NS fivebrane.  Thus, the moduli
of figure \ref{fig:paradox}(c) are the same as those of
figure \ref{fig:paradox}(a) or figure \ref{fig:paradox}(b).

The virtue of this proposal is that in figure \ref{fig:paradox}(c), unlike
figure \ref{fig:paradox}(b),
there is a mechanism to get a massless hypermultiplet whenever
$\vec x=\vec m$.  Precisely under this condition, the two threebranes
in figure \ref{fig:paradox}(c) meet in spacetime, and an elementary string
stretched between them should plausibly give a massless hypermultiplet.

In fact, the mechanism for getting a massless hypermultiplet in
figure \ref{fig:paradox}(a) is a special case of the mechanism in
figure \ref{fig:NS5D3}(b), while
the mechanism for getting a massless hypermultiplet in
figure \ref{fig:paradox}(c)
is a special case of the mechanism in figure \ref{fig:NS}.  So as promised
in section two, we have obtained (if our resolution of the paradox
can be justified) a phase transition in which one mechanism for generating
a massless hypermultiplet is converted to another.

\bigskip\noindent
{\it A Priori Argument}

Now we will give an {\it a priori} argument that the phenomenon
invoked in our resolution of the paradox must actually occur.

Let $X_{NS}$ be the world-volume of an NS fivebrane, and let
$X_D$ be the world-volume of a D fivebrane.
First of all, the NS fivebrane is the source of a three-form
field strength $H_{NS}$ which has the property that if $S$ is a small
three-sphere wrapping once around $X_{NS}$, then
\beq
\int_S{H_{NS}\over 2\pi}=1.
\label{nsb}
\eeq
Likewise, the D fivebrane is the source of a three-form field strength
$H_D$ such that if $S'$ is a small three-sphere wrapping
once around $X_D$, then
\beq
\int_{S'} {H_D\over 2\pi}=1.
\label{db}
\eeq

Let us consider Type IIB superstring theory on $\R^3\times M_7$,
with $M_7$ some seven-manifold.
Consider the case that $X_{NS}=\R^3\times Y_{NS}$ and
$X_D=\R^3\times Y_D$, where $Y_{NS}$ and $Y_D$ are three-manifolds
in $M_7$.  In our actual application $M_7=\R^7$, and $Y_{NS}$ (parametrized
by $x^3,x^4,x^5$) and $Y_D$ (parametrized by $x^7,x^8,x^9$)
are copies of $\R^3\subset \R^7$.
But temporarily we do not specify $M_7$, and we take
$Y_{NS}$ and $Y_D$ to be
compact and disjoint three-manifolds in $M_7$.   In this situation,
under a mild topological restriction ($Y_{NS}$ and $Y_D$ should
be trivial in $H_3(M_7,\Z)$) one can define the ``linking number'' of
the three-manifolds $Y_{NS}$ and $Y_D$ in the seven-manifold $M_7$.
Let $H_{NS}$ be the $H$-field created by the NS fivebrane on $\R^3\times
Y_{NS}$, and let $H_D$ be the $H$-field created by the D fivebrane on
$\R^3\times Y_D$.  The linking number is defined by
\beq
L(Y_{NS},Y_D)=\int_{Y_D}{H_{NS}\over 2\pi}
\label{gg}
\eeq
or equivalently by
\beq
L(Y_{NS},Y_D)=-\int_{Y_{NS}}{H_D\over 2\pi}.
\label{hh}
\eeq
(The integrals are taken at any fixed point $p\in \R^3$.)

Now locally one can  write $H_{NS}=dB_{NS}$ where $B_{NS}$ is a two-form
potential.  If $B_{NS}$ were gauge-invariant and hence globally defined,
it would follow that the integral in (\ref{gg}) would vanish.
In general, in Type IIB superstring theory, $B_{NS}$ is not gauge-invariant
and is only defined locally.  Something special happens, however,
on the world-volume of a D fivebrane.  On the fivebrane, there is a $U(1)$
gauge field $A_D$, of two-form field strength $F_D$, which ``mixes''
with the bulk two-form $B_{NS}$, in such a way that 
$\Lambda_{D}=B_{NS}-F_D$ is gauge invariant.  Since in the absence of
additional branes
\beq
dF_D=0,
\label{kk}
\eeq
we have when no other branes are present a global representation of $H_{NS}$
restricted to $Y_D$ as the exterior derivative of a gauge-invariant
field, namely
\beq
H_{NS}|_{Y_D}=d\Lambda_D.
\label{mm}
\eeq
When this is inserted in (\ref{gg}), we seem to learn that the linking
number $L(Y_{NS},Y_D)$ must vanish.  To be more precise, since what appears
in (\ref{mm}) is the {\it total} $H_{NS}$ due to all sources, this
argument would show the total linking number of $Y_D$ with
respect to all NS fivebranes adds up to zero.

If we start with a situation with just one NS fivebrane and one D fivebrane
of linking number zero, this would seem to imply that they cannot
``pass through each other'' and change the linking number. That would
mean that figure \ref{fig:paradox}(a) can be converted neither to
figure \ref{fig:paradox}(b) nor to figure \ref{fig:paradox}(c).
The two types of fivebrane are infinitely strong
and cannot interpenetrate!  Rather than this  bizarre interpretation,
there is a more straightforward possibility.  Consider adding a threebrane
whose world-volume is $\R^3\times C$, where $C$ is a curve in $M_7$,
and suppose that the threebrane ends on $X_D$, that is that $C$ ends
on the three-manifold $Y_D$.  Let $p$ be the point on $Y_D$ at which $C$ ends.
The boundary of a threebrane looks like a magnetic source on 
$X_D$, in the sense that (\ref{kk}) must be modified to read
\beq
dF_D = \pm \delta(p) 
\label{imbo}
\eeq
where the sign depends on the orientation, that is, on whether $C$ 
``begins'' or ``ends'' on $Y_D$.   (This basic fact, that a threebrane
boundary looks like a magnetic charge on the fivebrane, was of course
one of the starting points in section two.)
When this effect is included, the relation (\ref{mm}) implies not
that $H_{NS}|_{Y_D}=d\Lambda_D$, but that 
\beq
H_{NS}|_{Y_D}=d\Lambda_D-\sum_\alpha\epsilon_\alpha
\delta(p_\alpha),
\label{gjk}
\eeq
 where the $p_\alpha$ are points at which threebranes
end, and $\epsilon_\alpha=\pm 1$ are the corresponding signs.
When this equation is put in formula (\ref{gg}) for the linking
number, we learn not that the linking number vanishes; rather,
what vanishes is the sum of the linking number of $Y_D$ (with respect
to all NS fivebranes) plus the net number of threebranes ending on $Y_D$.

Therefore, when an NS fivebrane passes through a D fivebrane (so that
the linking number changes) a threebrane connecting them is created.
This is the claim that was made above in figure \ref{fig:paradox}(c).

By now we have explained the basic phenomenon, but we still
need to adapt the discussion to the noncompact situation
in which $M_7=\R_7$, and $Y_{NS}$ and $Y_D$ are both copies of
$\R^3$.  The main change is that the integral 
$\int_{Y_D}d\Lambda_D$ need no longer vanish; rather, it is determined
by the behavior of $\Lambda_D$ near infinity, and is a ``total magnetic
charge'' as seen by the fivebrane observer.  This total magnetic
charge need not vanish (it certainly did not vanish in the models considered
in section three!)  but because it can be measured at infinity it
is  conserved, and unchanged when we move fivebranes around.

So in our problems, to each fivebrane $X_{NS}$ or $X_D$
we will assign a ``total magnetic
charge,'' which will be conserved in all phase transitions.
The total magnetic charge on a given fivebrane, say $X_D$, is the 
sum of the linking numbers of $X_D$ with respect to all NS fivebranes,
plus the net number of threebranes ending on $X_D$.
However, because of the noncompactness, we need some care in defining
what we mean by the linking numbers.
Let us go back to figure \ref{fig:paradox}(a).  We let $X_{NS}$ be the
world-volume
of the NS fivebrane on the right, and $X_D$ be the world-volume of the 
D fivebrane.  We want to know their linking numbers, defined by a flux
integral as in (\ref{gg}) or (\ref{hh}) above.  Instead of actually evaluating 
integrals, we can note the following two facts.
\begin{enumerate}
\item By symmetry the linking integral
changes sign when the relative position of the two branes in the
$x^6$ direction is reversed. \item  It changes by $+1$ when they pass
through each other. 
\end{enumerate}
 These properties imply that the linking numbers,
defined by the flux integrals, are $\pm 1/2$ depending on which
brane is on the left.  (The integrals (\ref{gg}) and (\ref{hh}) would
always give integers when the manifolds involved are compact,
but not in the noncompact situation considered here.)

The signs can be stated precisely as follows.
Let $X_{NS}$ be a particular NS fivebrane.  Let $r$ be the
number of D fivebranes to the right of $X_{NS}$ and  $l$ the
number to the left.  Let $R$ be the number of threebranes which end
to the right \footnote{To avoid any 
possible confusion, if $X_{NS}$ is at $x^6=t$, then a threebrane
whose world-volume has $x^6>t$ ends to the right of $X_{NS}$ and one
whose world-volume has $x^6<t$ ends to the left.} of $X_{NS}$
 and $L$ the number of threebranes that end
to the left of $X_{NS}$.  Then the total magnetic charge measured on
$X_{NS}$ is
\beq
L_{NS}={1\over 2}(r-l)+(L-R).
\label{linkingNS}
\eeq
The same formula holds for a D fivebrane $X_D$; if $r$ and $l$ are now
the numbers of NS fivebranes on the right and left and $R$ and $L$ are
the numbers of threebranes ending to the right and the left, then
\beq
L_D={1\over 2}(r-l) +(L-R).
\label{linkingD}
\eeq
We will also call the total magnetic charge measured on a fivebrane
the total linking number of that brane (with respect to all other
fivebranes and threebranes).
{}From the above formulas, it follows that the total linking number,
summed over all branes, is zero.   
This is the only restriction on the linking numbers for a configuration
to exist, although requiring unbroken supersymmetry can impose
further restrictions.

As we will see, this process of creation of a threebrane by moving  fivebranes
through each other has far-reaching consequences for the three-dimensional
dynamics. 


There is an analogous process in $M$-theory, in which two
fivebranes crossing each other become connected by a twobrane.
It can be deduced in a similar fashion.

\subsection{$U(1)$ With One Electron}\label{sec:Uoneone}

As an ilustration of this  effect of threebrane creation, we
consider the simplest
configuration in which it happens.
In this example we take the numbers of NS and D fivebranes
to be $n_s=2, n_d=1$ and look at the possible theories which
emerge from this configuration. 
The parameters in the theory are two solitonic
five brane locations $\vec w_1, \vec w_2$, their singlet superpartners
$t_1, t_2$ and the D5 brane locations $\vec m$ and $z$. These parameters are
thought of as background fields (or coupling constants) for the three
dimensional gauge theory.
We can set the sum of the parameters for the solitonic five branes to
zero $\vec w_1=-\vec w_2=\vec w$, $t_1=-t_2=t$ by a choice of origin.
Similarly we can take the position of the D5 brane to be at the origin
$\vec m=0$.
We can 
not fix $z$ since the origin in the $x^6$ direction was already chosen.

We complete the specification of the model by saying that the 
total magnetic charges or linking numbers are to be as follows:
$-1/2$ and $1/2$ for the two NS fivebranes, and 0 for the D fivebrane.

\begin{figure}[htb]
\centerline{\psfig{figure=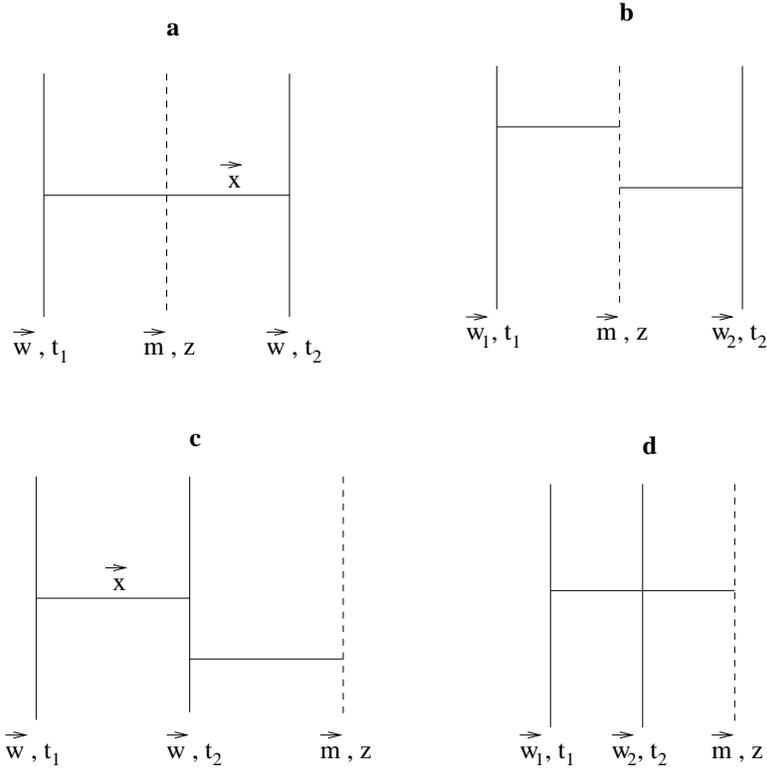,width=4in}}
\caption[]{$U(1)$ gauge theory with one electron.  In (a) and (c),
$\vec w_1=\vec w_2$, but not in (b) and (d), where an FI coupling
has been turned on.
\protect\label{fig:Uoneone}}
\end{figure}

We have two phases to consider, depending on the arrangement 
of the fivebranes. First let us take the D5 brane to
be between the two solitonic five branes, in the sense that
$t_1<z<t_2$.
The behavior further depends on whether (a)
  $\vec w=0$ or (b) 
 $\vec w\not= 0$.
 In case (a), a configuration with the promised
linking numbers can be built as in figure \ref{fig:Uoneone}(a), with
a threebrane stretched between the two fivebranes.  According
to our rules in section three, the effective gauge theory in
$2+1$ dimensions is then a $U(1)$ gauge theory with one charged
hypermultiplet.  Moreover, we learned in section three that the
Fayet-Iliopoulos $D$-term coupling is $\vec w$.  The transverse position
or $\vec x$ value of the threebrane in figure \ref{fig:Uoneone}(a)
parametrizes the Coulomb
branch (together with the dual of the photon).

Now we consider case (b), that is $\vec w\not= 0$.  
The field theory of $U(1)$ with one hypermultiplet and a $D$-term
exhibits a Higgs mechanism; there is a unique supersymmetric
vacuum, with a mass gap.  Let us see how this occurs in the present
context.  For $\vec w\not=0$, we cannot in a supersymmetric fashion
suspend a threebrane between the two fivebranes.  However, we can
build the 
configuration of \ref{fig:Uoneone}(b), which has the same linking
numbers.  In this configuration, there are no moduli, since all
threebranes connect fivebranes of opposite type.  So this agrees
with the field theory result that in the presence of the FI 
coupling
 the vacuum
is unique.  Note that to make an actual transition from \ref{fig:Uoneone}(a)
to \ref{fig:Uoneone}(b), the threebrane of figure \ref{fig:Uoneone}(a) must meet 
the 
D fivebrane in spacetime, which occurs only at $\vec x=0$, that
is, at the origin of the Coulomb branch.  This again agrees with
the field theory result.

In either case the $U(1)$ gauge coupling is $g^2=1/|t_2-t_1|$.  More
puzzling is the interpretation of the parameter $z$ in the low
energy theory.  There is no obvious dependence of the low energy
physics on $z$.  What happens if we increase $z$ so that $z>t_2$
(or equivalently reduce $z$ so that $z<t_1$)?

For $t_1<t_2<z$, the supersymmetric configurations with
the given linking numbers are sketched in figure \ref{fig:Uoneone}(c,d).
The case of $\vec w=0$ is in figure \ref{fig:Uoneone}(c).  Here we see
the same spectrum as in figure \ref{fig:Uoneone}(a): a $U(1)$ vector multiplet
with one charged hypermultiplet; its mass parameter is still
$\vec x$ so a singularity would have to be at $\vec x = 0$.
The case of $\vec w\not= 0$
is in figure \ref{fig:Uoneone}(d).  There is a single threebrane that stretches
between fivebranes of opposite type, and so supports no massless
modes or moduli.  This again agrees with figure \ref{fig:Uoneone}(b).

In this model, the phase structure seems to be entirely independent
of $z$ (though we cannot exclude the possibility that something
exotic may happen at $z=t_2$ or $z=t_1$, perhaps if also
$\vec x=0$).  The parameter $z$ appears to be an irrelevant, though
mysterious, perturbation.  

It is amusing to write the metric on the Coulomb branch (which is
a smooth Taub-NUT metric) in these variables.
It is
\beqar
ds^2&=&G(\vec x)d\vec x\cdot d\vec x+G(\vec x)^{-1}d\tilde\theta
d\tilde\theta,\qquad
d\tilde\theta=d\theta-\vec \omega\cdot d\vec x, \nonumber\cr
G(\vec x)&=&|t_1-t_2|+{1\over |\vec x|},\qquad
\nabla_a\times\vec\omega=\nabla_a({1\over|\vec x|}) .
\eeqar
This holds for any values of $t_1, t_2, z$.

\subsection{A Conundrum}

\begin{figure}[htb]
\centerline{\psfig{figure=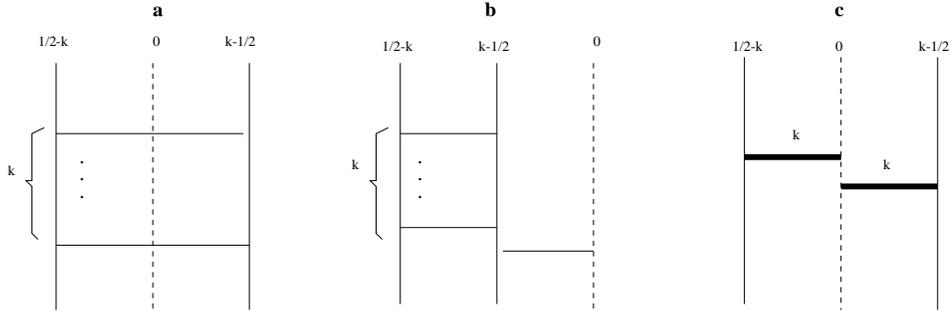,width=5in,angle=-90}}
\caption[]{$U(k)$ gauge theory with one flavor and its various phases.
Figure (a) 
and figure (b) have the same matter content and therefore are in the
same phase. The vertical solid lines are solitonic fivebranes and
the vertical dashed line is a fivebrane. The horizontal lines in (a)
are threebranes, two of which are shown explicitly.  In (c), 
the big fat line with a number $k$ attached to it represents $k$ threebranes
which stretch between an NS fivebrane and a D fivebrane.
The numbers on top of the five branes denote the linking numbers.  
\protect\label{fig:Uk1}}
\end{figure}

Next we can consider a model in which  there are still two NS fivebranes
and one D fivebrane, but we consider an arbitrary number $k$ 
of threebranes, as in figure \ref{fig:Uk1}(a).  To fully specify
the model, we must give 
the linking numbers, which are $\pm(\half - k)$ for
the NS fivebranes, and zero for the D fivebrane.  One reason to consider
this model in some detail is that it is the simplest case in which we meet
a certain puzzle.

Applying our usual rules to figure \ref{fig:Uk1}(a),
this theory can be identified as a $U(k)$ gauge theory with one flavor in the
fundamental representation. The gauge coupling is $|t_1-t_2|$, The 
Fayet-Iliopoulos coupling is
$\vec w_1-\vec w_2$.  First we consider the case where
this vanishes.  The bare mass term for the hypermultiplet
 is $\vec m$, and can be set to zero by shifting the coordinates.
The Colomb branch has quaternionic dimension $k$.
The model has no Higgs branch.
As in the $U(1)$ case we discussed before, 
the string theory description has a mysterious parameter
$z$ with no obvious meaning in the low energy physics.  
Let us see what happens when we increase $z$ and reach the regime
$t_1<t_2<z$.
Again a threebrane is generated when the fivebranes cross, to give the
picture of figure \ref{fig:Uk1}(b).  Using
our rules for matter fields we see that the
theory remains with the same matter content.
Again this flow is apparently irrelevant for this theory.

Now we try to turn on the Fayet-Iliopoulos coupling, that is
to take $\vec w_1-\vec w_2\not=0$.  
What happens in field theory in the present model is that this
triggers supersymmetry breaking.  
To see this, recall that a hypermultiplet
in the fundamental repreentation of $U(k)$ is equivalent from the
point of view of $N=2$ supersymmetry to a pair of multiplets
$A^i$ and $B_j$ in the fundamental representation and its dual (or
complex conjugate).
In the presence of the Fayet-Iliopoulos coupling, if the $N=2$ subalgebra
is chosen suitably, the conditions for unbroken supersymmetry become
\beqar
\eqalign{ A^iB_j & = 0 \cr
        A^i\bar A_j-B^i\bar B_j & = \delta^i{}_j w,\cr}
\label{deqns}
\eeqar
with $w$ the FI coupling.  For $k>1$, these equations have no solution.

What happens when we embed this question in string theory, displacing
the NS fivebranes to $\vec w_1\not= \vec w_2$?
A supersymmetric configuration
now cannot have a threebrane stretching between the two NS fivebranes.
There are two cases to consider: (i) $t_1<z<t_2$; (ii) $t_1<t_2<z$.

In case (i), to make a supersymmetric configuration 
all threebranes must be broken into
threebranes that go ``half way,'' starting on the left  NS fivebrane
and ending on the D fivebrane, or starting on the D fivebrane and 
ending on the right NS fivebrane.  This is sketched in figure \ref{fig:Uk1}(c).
No moduli are possible in this picture, as threebranes starting and ending
on fivebranes of opposite type are ``frozen.''

In case (ii), no supersymmetric configurations are possible at all.
The only threebrane configuration with $t_1<t_2<z$
and linking numbers $\half-k,k-\half$, and 0 is the one already
sketched in figure \ref{fig:Uk1}(b), and this is not compatible with
supersymmetry when $\vec w\not= 0$.

Thus, in region (ii), the string theory agrees with the field theory
result: supersymmetry is broken when the FI coupling is turned on.
In region (i), the string theory appears to give an isolated supersymmetric
vacuum, contradicting the field theory.

In general, we can see two possible resolutions of this puzzle:

(1) The parameter $z $ is not really irrelevant in the low energy
$2+1$ dimensional physics.  The classical equations
(\ref{deqns}) are valid when $t_2<z$ or $z<t_1$, but not when $t_1<z<t_2$,
and in that latter range there is a supersymmetric vacuum in the presence
of the FI terms.  

(2) A configuration (like that of figure \ref{fig:Uk1}(c) )
with more than one threebrane connecting
a given NS fivebrane to a given D fivebrane does not have a quantum
state of unbroken supersymmetry, though merely drawing the picture
suggests that it does.

Later, we will refer to a configuration with more than one threebrane
connecting an NS fivebrane to a D fivebrane as an $s$-configuration,
so the question is whether $s$-configurations can be supersymmetric.

As the above statement indicates, the physics in this model
is less exotic if $s$-configurations are not supersymmetric.
That is a general pattern that we will see also in other examples.
Note that starting with a configuration that is not an $s$-configuration,
crossing of fivebranes does not generate an $s$-configuration, so
it is consistent to assume that $s$-configurations are not supersymmetric.
However, we do not know a direct argument for this.

\begin{figure}[htb]
\centerline{\psfig{figure=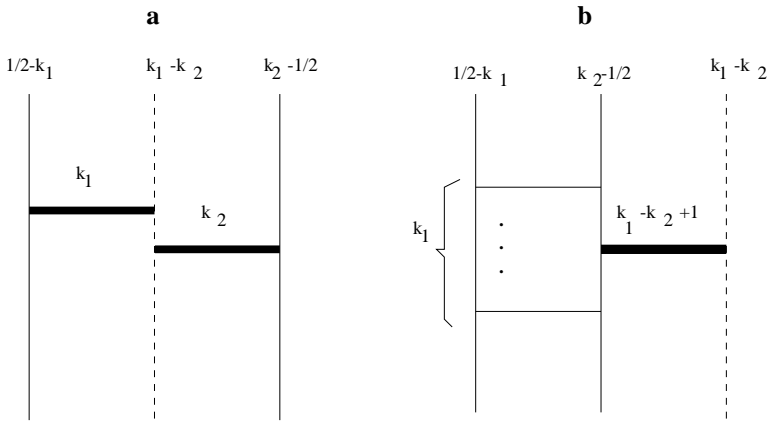,width=4in}}
\caption[]{Exotic phases which emerge 
in a configuration of two solitonic five
branes and a D5 brane if $s$-configurations are supersymmetric. 
The vertical solid lines are solitonic five branes and
the vertical 
dashed line is a D5 brane. The horizontal lines are threebranes.
The big fat line with a number $k$ attached to it represents $k$ 
threebranes
which stretch between a D5 brane and a solitonic five brane. 
The numbers on top of the five branes denote the linking numbers.
\protect\label{fig:konektwo}}
\end{figure}

We illustrate in  figure \ref{fig:konektwo}
some of the simplest of the bizarre possibilities that can arise
if $s$-configurations are   supersymmetric.  In part (b), we show
a $U(k_1)$ gauge theory coupled to an $s$-configuration labeled
by a positive integer; by turning
on an FI coupling, this can make a transition to the double   
$s$-configuration shown in (a).

\section{Mirror Symmetry At Work}

In this section we look at the general class of theories
labeled by $n_d$ D fivebranes and $n_s$ NS fivebranes, with different
numbers of threebranes connecting them.
The number of couplings (or background fields) 
of the theory is $4(n_d+n_s)-7$.
These are given by four transverse coordinates per  brane.
We are free to set 
the origin for all seven coordinates which are transverse to
the $0,1,2$ plane and this gives the above formula.

As one moves in moduli space, various phase transitions can occur:
\begin {enumerate}
\item Threebranes can reconnect when possible.
An example for that we saw in the case of $U(1)$ with one 
flavor in section
\ref{sec:Uoneone}. 
\item Fivebranes can be rearranged, passing through each other
and creating new threebranes, as in the previous section.
\end{enumerate}

The only invariants under all these transitions
are the ``linking numbers'' (integer or half integer) seen
by each fivebrane.   So models (but not phases) are classified 
by $n_d, n_s,$ and the linking numbers.  

Mirror symmetry is manifest in this class of models.
It exchanges the $3-4-5$ directions in spacetime with 
the $7-8-9$, directions, so it exchanges the Fayet-Iliopolous parameters
$\vec w$ with the bare masses $\vec m$.  It also
exchanges the two types of fivebrane, so it 
exchanges $n_d$ and $n_s$ while exchanging the two sets of linking numbers.

\subsection {$U(1)$ Gauge Theory With Two Flavors}

For our first experience with mirror symmetry, we consider
a model with $n_s=n_d=2$, and all linking numbers zero. 
This model is self-mirror because $n_s=n_d$ and all linking
numbers are equal.
\begin{figure}[htb]
\centerline{\psfig{figure=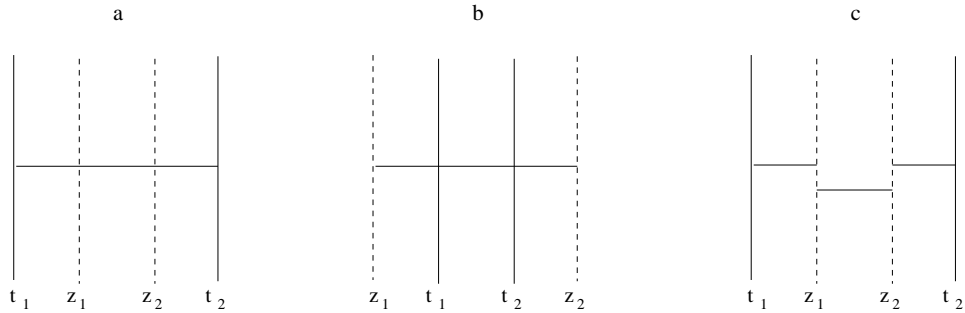,width=5in,angle=-90}}
\caption[]{$U(1)$ gauge theory with two electrons.
\protect\label{fig:Uonetwo}}
\end{figure}

In one obvious phase with two NS fivebranes on the outside connected
by one threebrane (figure \ref{fig:Uonetwo}(a)), this
is a $U(1)$ gauge theory  with two charged hypermultiplets.  
The self-mirror property of this model was already observed in
\cite{SI}, and in the present context is a consequence
of the general mirror symmetry of this class of models.
Mirror symmetry applied to the configuration of figure \ref{fig:Uonetwo}(a)
maps it to the configuration of figure \ref{fig:Uonetwo}(b), which by going 
through
a series of phase transitions can be mapped back to the configuration
of figure \ref{fig:Uonetwo}(a).  To be precise about these phase transitions,
one first moves the NS fivebranes back to the outside; some new
threebranes appear, as sketched in figure \ref{fig:Uonetwo}(c).  Then, by 
reconnecting
the threebranes to a single threebrane stretching between the two
NS fivebranes (after adjusting $\vec w_1-\vec w_2$ to zero to make
this possible), one gets back to the configuration of
figure \ref{fig:Uonetwo}(a).
The fact that this succession of operations gets us back to the starting
point is guaranteed by the fact that the linking numbers -- which uniquely
determine the number of threebranes between each pair of successive
fivebranes -- were chosen to be self-mirror.

\begin{figure}[htb]
\centerline{\psfig{figure=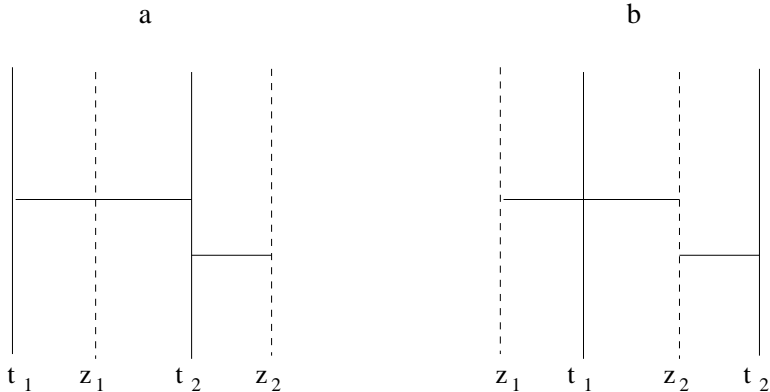,width=4in}}
\caption[]{$U(1)$ gauge theory with two electrons for a different
arrrangement of fivebranes.
\protect\label{fig:DNSDNS}}
\end{figure}

It is interesting to move the fivebranes around and ask what the model
looks like in other parts of the parameter space.  For example,
in  figures \ref{fig:DNSDNS}(a,b), we consider the situation in which 
$t_1<z_1<t_2<z_2$.
This configuration is mirror symmetric up to a rotation of, say,
the $x^1-x^6$ plane that reverses
the sign of $x^6$.  (This rotation also changes the sign of $x^1$
and so looks like a parity transformation in the low energy $2+1$
dimensional theory.)  With the threebranes connected as in
figure \ref{fig:DNSDNS}(a), this looks like an electric $U(1)$ gauge theory
with two charged hypermultiplets; with the threebranes connected
as in figure \ref{fig:DNSDNS}(b), it looks like a magnetic $U(1)$ gauge theory
with two charged hypermultiplets.  The transition between them is possible
only when $\vec w_1=\vec w_2$ and $\vec m_1=\vec m_2$, just as in
\cite{SI}.  

\begin{figure}[htb]
\centerline{\psfig{figure=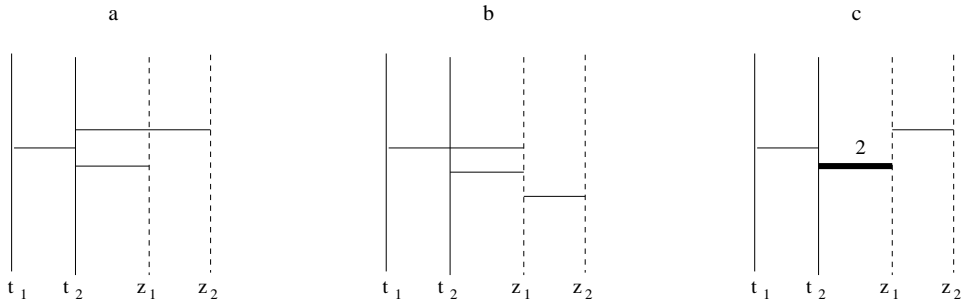,width=5in,angle=-90}}
\caption[]{$U(1)$ gauge theory with two electrons in the range
$t_1<t_2<z_1<z_2$.  Sketched
are the Coulomb branch, the Higgs branch, and an exotic branch (with
two threebranes connecting the adjacent NS and D fivebranes)
that exists if $s$-configurations are supersymmetric.
\protect\label{fig:newfig}}
\end{figure}

Starting with the configuration considered in the last paragraph, it
is interesting to increase $z_1$ until $t_2<z_1$ and both NS fivebranes
are to the left of both D fivebranes.  
A supersymmetric configuration in this situation is sketched in
figure \ref{fig:newfig}.  There is both a Coulomb branch and a Higgs branch, as 
indicated in the figure.  Applying the standard rules shows on these
branches the expected spectra: $U(1)$ with two charged hypermultiplets.
  Also shown in the figure is a more exotic situation, with
both a Higgs and a Coulomb modulus,
that is possible if $s$-configurations are allowed.

In all these figures, an overall motion of D fivebranes relative
to NS fivebranes is irrelevant for the $2+1$-dimensional
physics unless $s$-configurations are supersymmetric.  If
$s$-configurations are supersymmetric, the phase in figure
\ref{fig:newfig}(c)  is one that exists only for the given ordering
of branes, so its existence can be affected by adding a constant
to $z_1$ and $z_2$.  This illustrates the general fact that
one obtains simple and consistent results if one assumes
that $s$-configurations are not supersymmetric.


Now let us go back to the most tame situation of figure \ref{fig:Uonetwo}(a).
We interpret this as an electric $U(1)$ gauge theory with two charged
hypermultiplets whose bare mass parameters (being controlled by the
$\vec m$'s of the D fivebranes) are arbitrary and in particular can 
vanish.
When they vanish, a field theorist expects to see an enhanced $SU(2)_d$
global symmetry of the two hypermultiplets.  To get this in our
setup, we simply take the D fivebranes to be coincident, that is we take
$z_1=z_2$.  Then on the 
worldvolume of the D fivebranes there is an enhanced
$U(2)$ gauge symmetry, via the Chan-Paton factors, and after factoring
out the center of $U(2)$, which acts trivially on the threebranes,
this gives
 an $SU(2)$  global symmetry in the $2+1$-dimensional low energy theory.

On the other hand, if we would think in terms of the {\it magnetic}
gauge coupling, which is given by $g_m^{-2}=|z_1-z_2|$, it is clear
that this method of obtaining an enhanced $SU(2)_d$ global symmetry
involves setting $g_m=\infty$.  Conversely, with the electric gauge
coupling $g_e$ such that coupling  $g_e^{-2}=|t_1-t_2|$, we will
observe an enhanced $SU(2)_{ns}$ 
global symmetry exactly when $g_e=\infty$.
Comparing this to \cite{SI}, those authors discovered an enhanced
$SU(2)_{ns}$ 
global symmetry 
precisely when $g_e=\infty$.  In the conventional Lagrangian description,
there is no parameter analogous to $g_m$, and (bare masses being zero)
the $SU(2)_d$ symmetry is always present (this is the usual
flavor symmetry of two massless quarks).  Clearly, a major difference
between the brane configurations we are considering here and what
one can see in low energy field theory is that in the present description
the parameters $g_e$ and $g_m$ are simultaneously visible, though
their interpretations as gauge couplings hold in only part of the 
parameter space.  

Since in known Lagrangian field theory, one
is limited to $g_m=\infty$, mirror symmetry was seen in \cite{SI}
by setting also $g_e=\infty$.  In the present context, we see
mirror symmetry for finite $g_e$ and $g_m$.
The critical point of \cite{SI} with $g_e=g_m=\infty$ and 
$SU(2)_{ns}\times SU(2)_d$ symmetry corresponds to the maximally
symmetric situation with all four fivebranes at the same value of
$x^6$.  This configuration can be perturbed to a nine parameter
family, the parameters being three relative $x^6$ values,
three components of $\vec w_1-\vec w_2$, and three components
of $\vec m_1-\vec m_2$.  
(Assuming $s$-configurations are not supersymmetric, 
 one of the nine parameters, namely an overall shift in the $z_j$
for fixed $t_i$, is generally hard to see
 in the $2+1$-dimensional world; the
other eight parameters form two supermultiplets.) 
In any one of the Lagrangian realizations,
say via the electric gauge group, seven of the nine parameters
are visible ($t_1-t_2$, $\vec w_1-\vec w_2$, and $\vec m_1-\vec m_2$).
Some subloci of the nine parameter family are particularly difficult
to understand in the low energy field theory, namely those in
which a D and an NS fivebrane are still coincident.

Much of the physics of this model is controlled by
the  three relative distances along the $x^6$
direction.
We have the distance between the two solitonic five branes $|t_1-t_2|$,
which is an electric gauge coupling,
the distance between the two D5 branes $|z_1-z_2|$,
which is a magnetic gauge coupling and the distance between one
solitonic five brane and one D5 brane $|t_1-z_2|$, which is
more difficult to interpret.
When the first distance vanishes we get the usual non-linear sigma model
associated to hypermultiplet moduli spaces or instanton moduli spaces.
For this particular case the 
relevant moduli space is that of $SU(2)$ instantons.
When the second distance vanishes we get a flow of the gauge theory to a
strongly coupled theory.
When the third distance vanishes we get a  model that must have
peculiar properties, since at the ``bicritical point'' $z_1=t_1$,
$z_2=t_2$,  one has a model that 
can be perturbed to either an electric or magnetic gauge theory.

\bigskip\noindent
{\it Weak Coupling And Four Dimensions}

One of the most unusual features of our description, relative to conventional
three-dimensional field theory, is that it is possible to see at the same
time electric and magnetic gauge couplings.  

This is related to the fact that
our threebrane world-volumes are really four-dimensional, if one looks
at them closely.  For the long wavelength behavior, one always has an
effective three-dimensional theory, but modes that carry momentum in
the fourth dimension must be included if one wants to discuss the physics
at energies of order $g^2$.  The theories we are considering here
(1) have the same infrared behavior as the conventional three-dimensional
gauge theories, and (2) have better duality properties, in the sense
that various statements that in field theory are only true in the infrared
are here exact.

If one goes to the weak coupling limit for either electric or magnetic
couplings, two fivebranes become far apart, and the threebranes connecting
them become macroscopically 
four-dimensional.  An attempt to take a weak coupling
limit would restore the full four-dimensional symmetry, giving a continuous
mass spectrum from a three-dimensional point of view.  
The relevant four-dimensional theory has the full ${\cal N}=4$ supersymmetry,
broken down to ${\cal N}=2$ by boundary conditions at the ends of
threebranes.

\bigskip\noindent
{\it The  Moduli Space Of Vacua }

By the Coulomb branch of the moduli space of vacua of this
theory, we mean a branch that is parametrized by the position
of a threebrane that ends on NS fivebranes (plus the superpartner
of those variables).  This is a four-dimensional hyper-Kahler
manifold that exists only when $\vec w_1-\vec w_2=0$.  By the
Higgs branch of moduli space, we mean a branch that is parametrized
by the position of a threebrane that ends on D fivebranes (plus
superpartner).   This is a four-dimensional hyper-Kahler manifold
that exists only when $\vec m_1-\vec m_2=0$.

\begin{figure}[htb]
\centerline{\psfig{figure=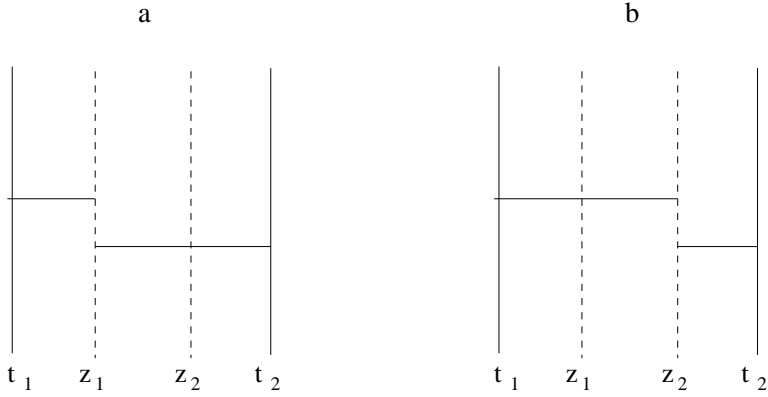,width=4in}}
\caption[]{The 
two possible vacua for $U(1)$ gauge theory with two electrons with
both bare mass and FI terms present.
\protect\label{fig:vacuatwocases}}
\end{figure}

When $\vec w_1-\vec w_2\not=0$ and $\vec m_1-\vec m_2\not=0$,
there is neither a Coulomb branch nor a Higgs branch.  In field
theory, in this situation, a standard analysis shows that there
are two isolated vacua, each with a mass gap.
This result is reproduced in the present context via
the pictures   sketched in figure \ref{fig:vacuatwocases}  for the
case $t_1<z_1<z_2<t_2$.  Similar pictures can be drawn for
other arrangements of fivebranes.

Now we will discuss the metric on the moduli space of vacua.

By looking at the system from a solitonic fivebrane point of view we
identify the Coulomb branch with the moduli 
space of an $SU(2)$ monopole in
the presence of two fixed monopoles represented by the mass deformation.
The overall center of mass of the two fixed monopoles can be absorbed in a
redefinition of the moduli $\vec x$.
The metric on the Coulomb branch of the moduli space has the form
\beqar
ds^2&=&G(\vec x)d\vec x\cdot d\vec x+G(\vec x)^{-1}d\tilde\theta
d\tilde\theta,\qquad
d\tilde\theta=d\theta-\vec \omega\cdot d\vec x, \nonumber\cr
G(\vec x)&=&|t_1-t_2|+
{1\over |\vec x+\vec m|}+{1\over |\vec x-\vec m|},\qquad
\nabla_a\times\vec\omega=\nabla_a({1\over|\vec x|}) . 
\eeqar
As for the metric on the Higgs branch, at $g_m=\infty$ it
can be determined by a classical calculation in the electric
gauge theory, and is equal to the flat metric on $\R^4/\Z_2$
(which is also an $SU(2)$ instanton moduli space on $\R^4$).
But in general we have $g_m\not= \infty$; the metric on the Higgs
branch is naturally described via the magnetic gauge theory
and is just isomorphic to what was written above:
\beqar
ds^2&=&G(\vec y)d\vec y\cdot d\vec y+G(\vec y)^{-1}d\tilde\theta
d\tilde\theta,\qquad
d\tilde\theta=d\theta-\vec \omega\cdot d\vec y, \nonumber\cr
G(\vec x)&=&|z_1-z_2|+
{1\over |\vec y+\vec w|}+{1\over |\vec y-\vec w|},\qquad
\nabla_b\times\vec\omega=\nabla_b({1\over|\vec y|}) .
\eeqar
This reduces to the flat metric on $\R^4/\Z_2$ when $z_1-z_2=0$
and $\vec w=0$.

The Fayet-Iliopoulus term can be written in the following form.
Denote the two hypermultiplets by $M_i,\tilde M_i,$ $i=1,2$. These fields are
$N=1$ $d=4$ chiral multiplets. The scalars in each of 
these chiral multiplets transform as $(2,1)$ of $SO(3)_H\times SO(3)_V$.
We collect them into a matrix form
\beq
M^{AA'}=\pmatrix{M&\tilde M^*\cr-\tilde M&M^*\cr}
\eeq
which satisfy the reality condition
\beq
M^*_{AA'}=\epsilon_{AB}\epsilon_{A'B'}M^{BB'}
\eeq
Then the condition for a supersymmetric vacuum is
\beq
\vec w={1\over2}\sum_i(M_i^*)_{AA'}\vec \sigma^{A'}_BM_i^{BA}
\eeq
Where $\vec \sigma$ are Pauli matrices.
We see that when $\vec m=\vec w=0$ the global 
$SU(2)_d\times SU(2)_H$ (the factors are respectively
the global symmetry present for  coincident D fivebranes
and the $R$ symmetry) is broken
spontanously on the Higgs branch to a diagonal subgroup.
In addition the scalars $z_i$ 
couple to the gauge field through the coupling
\beq
\int d^3x(A_\mu^i-\partial_\mu z_i)^2
\eeq

Mirror symmetry exchanges the two branches of the moduli space, the D
fivebrane position $\vec m$ with the solitonic fivebrane position
$\vec w$ and the gauge
coupling $t_i$ with the hidden gauge coupling for the Higgs branch $z_i$.
The magnetic  hypermultiplets 
$W_i,\tilde W_i,$ $i=1,2$ of the mirror theory can be treated
similarly. Each such multiplet transforms under
$(1,2)$ of $SO(3)_H\times SO(3)_V$.
We collect them into a matrix form
\beq
W^{AA'}=\pmatrix{W&\tilde W^*\cr-\tilde W&W^*\cr}
\eeq
which satisfy the reality condition
\beq
W^*_{AA'}=\epsilon_{AB}\epsilon_{A'B'}W^{BB'}
\eeq
Then the condition for a supersymmetric vacuum is
\beq
\vec m={1\over2}\sum_i(W_i^*)_{AA'}\vec \sigma^{A'}_BW_i^{BA}
\eeq
We see that when $\vec m=\vec w=0$ the global $SU(2)_s\times SU(2)_V$ is broken
spontanously on the Coulomb branch to a diagonal subgroup.
In addition the scalars $t_i$ couple to the gauge field through the coupling
\beq
\int d^3x(B_\mu^i-\partial_\mu t_i)^2
\eeq

\bigskip\noindent
{\it Higher Rank Gauge Groups}

Now we will briefly 
consider a larger class of models still with two NS fivebranes
and two D fivebranes but with more general linking numbers.
We start with a NS-D-D-NS configuration and assign linking
numbers $k_1, k_2, k_3, k_4$ to the five branes. 
These numbers are invariants
of the theories. We 
denote by $n_1, n_2, n_3$ the number of D3 branes between
each two adjacent five branes starting from the left.
Using equations \ref{linkingD} and \ref{linkingNS}
they are related to $k$'s by the formulas
\beq
k_1=1-n_1,\quad k_2=n_1-n_2,\quad k_3=n_2-n_3,\quad k_4=n_3-1.
\eeq
Note that the linking numbers sum to zero, reflecting the fact that
there are no sources for linking number at the boundary.
As
we reorder the fivebranes, the number of threebranes will change
in such a way that the linking numbers remain invariant.
Once we know these numbers we can use our rules in 
section \ref{sec:matter} to
find the spectrum in any particular phase.
The different phases and 
corresponding numbers are given in the table below.  The threebrane
numbers do not completely specify a branch of vacua, since one
must also specify how the threebranes are connected, as we discussed
in some detail above for $k_1=\dots = k_4=0$.

\begin{center}
\label{Rc}
\begin{tabular}{|c|c|c|c||c|c|c|}
\hline
NS & D & D & NS & $n_1$ & $n_2$ & $n_3$ \\ \hline
NS & D & NS & D & $n_1$ & $n_2$ & $n_2-n_3+1$ \\ \hline
NS & NS & D & D & $n_1$ & $n_1-n_3+2$ & $n_2-n_3+1$ \\ \hline
D & NS & NS & D & $n_2-n_1+1$ & $n_2$ & $n_2-n_3+1$ \\ \hline
\end{tabular}
\end{center}

Some additional orderings are related to these by obvious symmetries.
Only orderings in which the threebrane numbers are all non-negative
can give supersymmmetric vacua.

In most cases, for most values of the $n_i$, to analyze this
model one meets our basic question of whether $s$-configurations
are supersymmetric.

\subsection{Some Further Models}

Now we will analyze in some detail a model that is not self-mirror,
but has a known mirror. This is the model with
$n_s=2$ NS fivebranes and  $n_d=m$ D fivebrames.
It  is mirror to the case of $m$ NS and 2 D fivebranes
and gives, as we will see,
one of the examples of Intriligator and Seiberg \cite{SI}.

\begin{figure}[htb]
\centerline{\psfig{figure=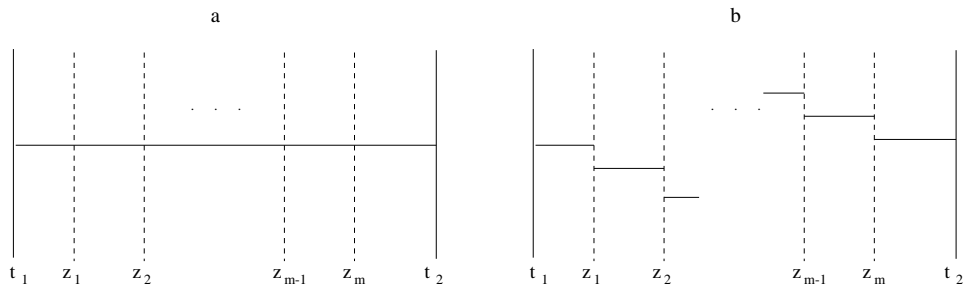,width=5in,angle=-90}}
\caption[]{$U(1)$ gauge theory with $m$ electrons.
\protect\label{fig:Uonen}}
\end{figure}

First we consider, as in figure \ref{fig:Uonen} the case that
 $t_1<z_1<\cdots,<z_m<t_2$.
We take the  linking numbers to be zero for the D fivebranes 
and $m/2-1$ for the left
solitonic fivebrane and $1-m/2$ for the right one.
For $\vec w_1=\vec w_2$,
there is a phase with one threebrane which stretches between the two
solitonic five branes.  In this phase, we can identify the theory
as a $U(1)$ gauge theory with $m$ charged hypermultiplets.

The Coulomb branch is four-dimensional and is parametrized as usual
by the scalars
(the position $\vec x$ and the scalar dual to the photon) associated with
the threebrane.
The hypermultiplet bare masses are 
given by the positions $\vec m_i$ of the D fivebranes.
The center of mass of the D fivebranes 
can be absorbed into a redefinition of the 
threebrane position and thus we have $m-1$ independent mass parameters. 
The difference $\vec w_1-\vec w_2$ is a Fayet-Iliopoulos coupling;
the Coulomb branch only exists if it vanishes.

With the same values of the couplings (such as $t_i$ and $z_j$),
there is also a Higgs branch in which the threebranes are
connected differently.  This is sketched in figure \ref{fig:Uonen}(b).
The Higgs branch is parametrized by $m-1$ threebranes 
which stretch between two
adjacent D fivebranes.  


\begin{figure}[htb]
\centerline{\psfig{figure=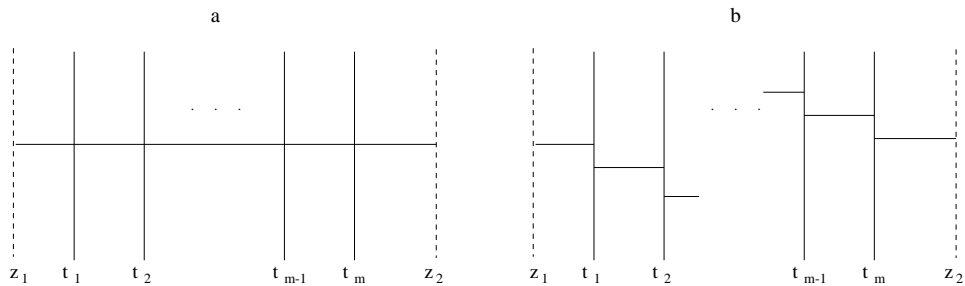,width=5in,angle=-90}}
\caption[]{$U(1)$ gauge theory with $m$ electrons in the mirror
description.
\protect\label{fig:dualUonen}}
\end{figure}

Now we apply mirror symmetry to the Higgs branch, which turns
into the configuration of figure \ref{fig:dualUonen} with two D fivebranes
on the outside and $m$ NS fivebranes on the inside.
In this case, we see a $U(1)^{m-1}$ gauge theory, 
one $U(1)$ factor for each threebrane connecting two adjacent
NS fivebranes.  Whenever two consecutive threebranes meet in
space one gets a massless hypermultiplet, so this theory has 
 $m$ hypermultiplets, of charges $(-1,0,\dots,0)$, $(1,-1,0,\ldots,0)$,
$(0,1,-1,0,\ldots,0)$, $\cdots$, $(0,\ldots,0,1,-1)$,
$(0,\dots,0,1)$.  So the $U(1)$ theory with $m$ hypermultiplets
must be mirror to the $U(1)^{m-1}$ theory with that particular
spectrum.  This is a result of Intriligator and Seiberg \cite{SI}.
As in their description, there is an $SU(m)$ global symmetry
when the $m$ D fivebranes, in the original description, are
coincident, and an $SU(2)$ global symmetry when the electric
gauge coupling of the original description is infinite, that is
when $t_1=t_2$. 

In our discussion of the Higgs branch, we  looked
at the maximum dimension component with the largest number of threebranes,
with threebranes connecting consecutive fivebranes.
It is possible to consider other cases.  In the description in terms
of the magnetic gauge group $U(1)^{m-1}$, these can be interpreted
in terms of partial Higgsing.  

\bigskip\noindent
{\it $U(2)$ With $m$ Flavors}

\begin{figure}[htb]
\centerline{\psfig{figure=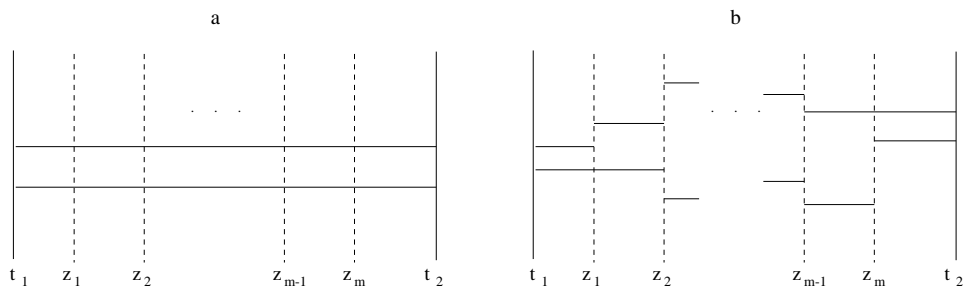,width=5in,angle=-90}}
\caption[]{$U(2)$ gauge theory with $m$ flavors. The Coulomb
and Higgs branches are sketched, respectively, in (a) and (b).
\protect\label{fig:2nflavor}}
\end{figure}

Next we consider a $U(2)$ gauge theory coupled to $m$ flavors.
This can be realized by the brane configuration of figure 
\ref{fig:2nflavor}. 

If we henceforth count dimensions in the quaternionic sense, then
the Coulomb branch is two-dimensional
and the Higgs branch is $2m-4$-dimensional.
We tune the masses to the origin in order to allow for 
threebranes to be created.
A Higgs phase is shown in figure \ref{fig:2nflavor}(b).
The magnetic gauge group is
is $U(1)\times U(2)^{m-3}\times U(1)$ with in an obvious notation 
hypermultiplets in the representation
${\bf (1,2)\oplus 2\oplus (2,2)\oplus
\ldots\oplus (2,2)\oplus 2\oplus (2,1)}$, 
coming from strings which stretch between
D threebranes along D fivebranes.
This system has a Coulomb branch of dimension $2m-4$ and a Higgs branch of
dimension $2$ which gives the required dimension for the dual theory.

This is a derivation of the mirror symmetry for a slightly different
model from the one found by \cite{SI}. In that example the gauge group is
$SU(2)$. How to compare the two models is discussed at the end of
this paper.

\subsection{More Mirrors}

\begin{figure}[htb]
\centerline{\psfig{figure=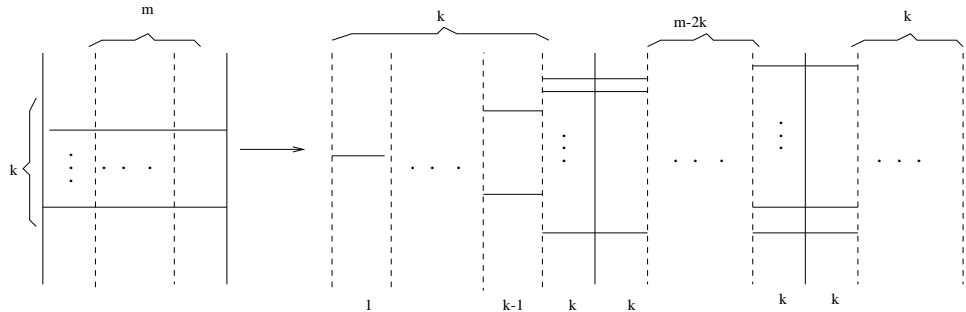,width=5in,angle=-90}}
\caption[]{A fivebrane configuration with $U(k)$ gauge group coupled to $m$
flavors, and its magnetic dual.  In figure (a) there 
are $m$ D fivebranes (dashed vertical lines) and $k$ threebranes
(horizontal lines) which stretch between two solitonic fivebranes
(solid vertical lines). This is the electric theory. 
In figure (b) we have the
magnetic dual. The numbers written
between a pair of adjacent fivebranes equals the number of 
threebranes stretched in between.
\protect\label{fig:Ukn}}
\end{figure}

In this section we will find mirrors for $U(k)$ gauge theories with
$m$ flavors for all $k$ and $m$, generalizing the above.
We first need to construct such a theory using branes. 
Using our rules from
section \ref{sec:matter} we find that a configuration of
two NS fivebranes and $m$ D fivebranes provides the right setting.
The five branes are ordered by  $t_1<z_1<\cdots<z_n<t_2$ as in
figure \ref{fig:Ukn}.

We start with an electric theory which has $k$ 
threebranes stretched between the
two solitonic five branes. It represents a gauge group $U(k)$.
The linking numbers are zero for the D5 branes, $m/2-k$ for the left
solitonic five brane and $k-m/2$ for the right one.
Note that for the special value $m=2k$ the 
linking numbers for the solitonic
fivebranes changes sign. This hints that 
the behavior of the theories for cases
above and below this critical value will be different.
That is actually so.  In fact, if the bare masses and FI term
are zero, complete Higgsing is only possible for $m\geq 2k$.
This is closely related to a result of \cite{APS}, 
where the $SU(k)$ theory with $m$ flavors was analyzed and it was
shown that for $m<2k$, complete Higgsing occurred only on a ``baryonic
branch.'' (The classical Higgs branch is
independent of the spacetime dimension in the range from three to six;
the work of \cite{APS} was motivated by the four-dimensional model.)
In our case, the gauge group is $U(k)$ instead of $SU(k)$; the gauging
of the extra $U(1)$ kills the baryonic branch, since the ``baryons''
are not $U(1)$-invariant.
It is presumably no coincidence that   in the $d=4$ theory,
the beta function of the $SU(k)$ part of the gauge group
changes sign for $m=2k$.
Going back to three dimensions, the fact that complete Higgsing
is only possible for $m\geq 2k$ means that a mirror -- understood
as a model whose Coulomb branch is the Higgs branch of the original
model -- can only exist in this range of $m$.

We will study the phases of the model by the usual techniques
of rearrangement of branes.
This produces a rich structure of phases
and provides a web of phase transitions; most of the phases (modulo
possible $s$-configurations) are mixtures of Coulomb and Higgs
branches, described in various ways.
To find a mirror,
 we must go to a Higgs branch in which all moduli are
derived from positions of threebranes that end on D fivebranes;
after this, we perform a mirror transformation and reinterpret the
moduli space as the Coulomb branch of a gauge theory.  Though the
answer we will get is complicated, it is obtained by  straightforward
implementation of the rules that we have described.
We want to find a mirror with zero bare masses and FI terms;
those can always then be added as perturbations.

There are many ways to go to a Higgs branch.
One approach is to
 move $k$ D fivebranes to the left and $k$ D fivebranes to the right.
Note that to do that we assume that $m\ge2k$.
The masses and FI terms are set to zero to allow 
for transitions which will generate D3 branes.
The fivebranes are ordered along the $x^6$ direction as
$z_1<\cdots z_k<t_1<z_{k+1}<\cdots<z_{m-k}<t_2<z_{m-k+1}<\cdots<z_m$.
The linking numbers are invariant 
under such reorderings and lead to generation
of threebranes for each reordering.
The linking numbers determine uniquely the number of threebranes 
between any two
fivebranes.  They are $1,2, \ldots, k, \ldots, k, k-1, \ldots, 1$.
We can now connect and break the threebranes in such a way that they will be
stretched between D fivebranes only.
(This can be done without ever passing through an $s$-configuration
as an intermediate state, so there is no need to assume they exist.)
After stretching all threebranes between D fivebranes as in the
figure, the  magnetic gauge theory is 
$U(1)\times U(2)\times\cdots\times U(k-1)\times U(k)^{m-2k+1}\times 
U(k-1)\times
\cdots\times U(1)$ with hypermultiplets transforming as
${\bf (1,2)\oplus
\ldots\oplus(k-1,\bar k)\oplus k\oplus
(k,\bar k)\oplus\ldots\oplus(k,\bar k)\oplus k\oplus (k,\overline{k-1})
\oplus \ldots\oplus (2,1)}.$
The dimension of the Coulomb branch of this magnetic theory
is calculated by counting the
number of threebranes and is $k(m-k)$, which equals the dimension of
the Higgs branch of the original electric theory; likewise,
 the magnetic theory has a Higgs branch of dimension $k$,
the dimension of the original Coulomb branch.

One could also have gone to the Higgs branch without moving fivebranes
at all, just by reconnecting threebranes.  For a Higgs branch,
all threebranes either (a) connect an NS fivebrane to a D  fivebrane
and so have no moduli, or (b) connect to D fivebranes to each other
and so represent electric hypermultiplets (or magnetic vectors).
The Higgs branch must then be a situation in which all $k$ threebranes
emanating from a NS fivebrane end on D fivebranes.  Assuming that
$s$-configurations are not supersymmetric, they must end on $k$
distinct D fivebranes.  For $m\geq 2k$, it is possible
to arrange so that each D fivebrane is connected by a threebrane to
at most one NS fivebrane; this will then give a point on a Higgs branch.
For $m<2k$, some D fivebrane is connected by threebranes to both
NS fivebranes.  Those threebranes could reconnect to give a threebrane
suspended between the two NS fivebranes, showing that the configuration
in question is on a branch with at least one massless vector
multiplet, and complete Higgsing has not occurred.  In this way,
we recover from the brane diagrams the field theory result that
complete Higgsing does not occur for $m<2k$.

For $m\geq 2k$, one can go on in this way and use the brane pictures
to compute the dimension of the Higgs branch.  For this, it is necessary
to make sure to use a brane picture which is generic (and does not
represent a sublocus of smaller dimension).  Generically, the $k$ 
threebranes connecting an NS fivebrane to D fivebranes should
connnect it to the closest ones (otherwise, these threebranes could
break); with this understood, the number of Higgs moduli -- interpreted
as the number of variable positions of threebranes that connect
D fivebranes -- is $ (m-1)k-2(1+2+\ldots +(k-1))=k(m-k)$,
in agreement with the field theory answer.  The description we have
given can actually be made more precise; the threebrane positions
correspond to the matrix elements of the Higgs matrices used in 
\cite{APS}.

\bigskip\noindent
{\it Continuation Past Infinite Coupling}

So far we have mainly considered transitions in which the ordering of
two fivebranes of opposite type is changed.  In such a transition,
a threebrane is created but (if $s$-configurations are not supersymmetric)
the field content of the low energy theory does not change.  Let us now
briefly consider the case of reordering two fivebranes of the same
type that have different linking numbers.
 
In this particular example, consider moving the rightmost NS fivebrane
to the extreme left and vice-versa.  If one simply keeps track of the linking
numbers, one sees that this process should turn a $U(k)$ theory with $m$
flavors into a $U(m-k)$ theory with $m$ flavors.
This is an attractive result, very reminiscent of Seiberg duality in 
four dimensions.  But there is obviously a problem if $m-k$ is negative. 

We propose that the resolution of this is that if the relative $D$-terms
vanish, then the two NS fivebranes, when they meet in $x^6$, actually
meet in spacetime and may exchange magnetic charge.  As a result, the
transition proposed in the last paragraph might not occur.

To avoid magnetic charge exchange, one can first turn on a $D$ term so
that the NS fivebranes can be reordered in $x^6$ without meeting in
spacetime.   We have already seen that turning on the $D$ term while
preserving supersymmetry requires $m\geq k$.  For this range of $m$ and $k$,
by first turning on the $D$ term, then moving the NS fivebranes, and then
turning the $D$ term back off, we can indeed interpolate between
the $U(k)$ and $U(m-k)$ theories with $m$ flavors.

For $m<k$, where one is restricted to $D=0$,
if one tries to reorder the two NS fivebranes, magnetic charge
exchange must occur when the fivebranes meet in spacetime, to avoid
the contradiction of a $U(m-k)$ gauge group with $m-k<0$.  For $m\geq k$,
the magnetic charge exchange does not occur for non-zero $D$ term, and
it is very plausible that it does not occur also when the $D$ term is zero.

If so, then the point at which the two NS fivebranes
meet in spacetime is a very interesting exotic fixed point.
It is a kind of common strong coupling limit of the $U(k)$ theory with
$m$ flavors and the $U(m-k)$ theory with $m$ flavors.
{}From this point of view, either 
one of these  theories is the analytic continuation
of the other to negative values of $1/g^2$.
   
\bigskip\noindent{\it Mirror For $SU(k)$}

It is also interesting
to obtain a mirror for
 a $SU(k)$ gauge theory with
$m$ flavors. This is formally obtained 
from the $U(k)$ theory that we have already studied
by ungauging the $U(1)$ which is the center of $U(k)$.
Ungauging the $U(1)$ reduces the dimension of the Coulomb branch
by one and increases the dimension of the Higgs branch by one.
We should look for an operation on the mirror theory that has
the opposite effect: increasing by one the dimension of the Coulomb
branch and reducing by one the dimension of the Higgs branch.
We conjecture that the mirror of ungauging a $U(1)$ is gauging
a $U(1)$, which has this effect.  

Here is a heuristic reason
for the conjecture to be true.  Consider any gauge theory $T$ with
a $U(1)$ gauge field $A$ with gauge transformation law $A_i\to
A_i-\partial_i a$, $a$ being the gauge parameter.  
Ungauging can be accomplished by adding
another hypermultiplet $H$ with a gauge transformation law $H\to
H+\epsilon a$, where $\epsilon$ is an arbitrary constant.  
No matter how small $\epsilon$ is, $H$ can be  gauged away, the
$U(1)$ gets a mass and can be integrated out, and the theory $T'$
with the $H$ field is equivalent at long distances to the $T$ theory
with the $U(1)$ ungauged.
Now, let us look for a mirror of $T'$. 
Because $\epsilon$ can be arbitrarily small, $H$ can be   treated
as a spectator, and if a mirror $\tilde T$ is known for $T$, a
mirror $\tilde T'$ can be found for $T'$ just by performing
the $T\to \tilde T$ mirror transformation in the presence of
$H$.  Under this transformation, $H$, being a hypermultiplet,
will very plausibly be reinterpreted as a $U(1)$ vector multiplet,
and if so the $\tilde T'$ theory will be $\tilde T$ with gauging
of an extra $U(1)$.  (It is interesting to note that such gauging
of $U(1)$'s that generate translations as well as rotations 
is important in constructions of some of the hyper-Kahler
manifolds we met in section five as hyper-Kahler quotients of
Euclidean space \cite{rychenkova}.) 

\begin{figure}[htb]
\centerline{\psfig{figure=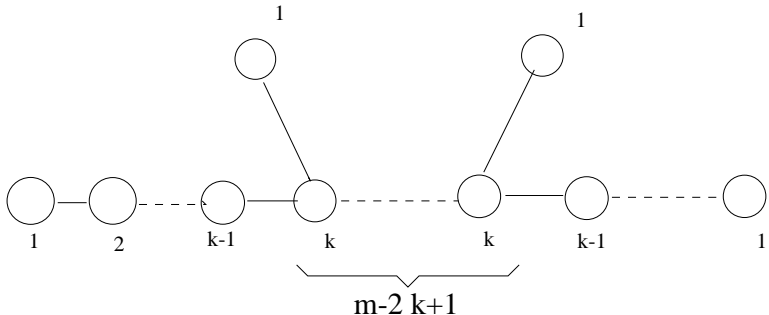,width=4in}}
\caption[]{A Dynkin 
diagram which represents the magnetic theory which is dual
to $SU(k)$ with $m$ flavors. The nodes represent factors of the
gauge group and the lines connecting nodes represent hypermultiplets.
\protect\label{fig:SUkn}}
\end{figure}

Going back to our particular problem,
it has two obvious global $U(1)$ symmetries,
which come from the $U(1)$ gauge symmetries on the NS fivebrane
worldvolumes, interpreted as global symmetries in the $2+1$-dimensional
world.  We want to gauge a combination of the two $U(1)$'s,
but as the diagonal $U(1)$ (which gauges overall translations)
decouples anyway, there is no harm in gauging both of these $U(1)$'s.
The magnetic dual we obtained above, once one gauges
  both of the extra $U(1)$'s, can be described by the quiver diagram
of figure \ref{fig:SUkn}.
In a quiver diagram, each node labeled by an integer $a$ represents
a $U(a)$ factor in the gauge group, and each line between  two
nodes labeled by $a,b$ represents a hypermultiplet transforming
as ${\bf (a,\bar b)}$ under $U(a)\times U(b)$.  Note that the
particular quiver diagram in figure \ref{fig:SUkn}, in the case $k=2$,
is the quiver associated with the $D_n$  extended Dynkin diagram.   
Our assertion that the theory associated with the quiver in the 
figure represents the mirror of  the $SU(k)$ theory with 
$m$ hypermultiplets reduces for $k=2$ to an assertion by
Intriligator and Seiberg \cite{SI}.

\section*{Acknowledgements}
We would like to acknowledge helpful discussions with G. Chalmers, 
O. Ganor, V. Sadov,
N. Seiberg, and M. Strassler.  While this work was nearing completion
two additional papers explaining three-dimensional mirror symmetry
in some situations appeared \cite{deBoer:1996,Zaffaroni}.   
\bibliography{SLtwo}
\bibliographystyle{utphys}

\end{document}